\documentclass[12pt]{article}
\usepackage[a4paper,margin=25mm]{geometry}
\usepackage[T1]{fontenc}
\usepackage[utf8]{inputenc}
\usepackage{lmodern}
\usepackage{setspace}
\usepackage{amsmath,amssymb,mathtools,amsthm}
\allowdisplaybreaks[3]
\usepackage{booktabs,array}
\usepackage{graphicx}
\usepackage{tikz}
\usetikzlibrary{positioning}
\usepackage{microtype}
\usepackage{placeins}
\usepackage[authoryear,round]{natbib}
\usepackage[colorlinks=true,linkcolor=blue,citecolor=blue,urlcolor=blue]{hyperref}
\hypersetup{
  pdfauthor={Kiyoshi Asai},
  pdftitle={From Derivatives to Exact Sequence Substitution Effects in Dynamic Programming for Biological Sequence Analysis},
}
\theoremstyle{plain}
\newtheorem{proposition}{Proposition}

\title{From Derivatives to Exact Sequence Substitution Effects in
Dynamic Programming for Biological Sequence Analysis}
\author{%
  Kiyoshi Asai$^{1,2}$\thanks{Correspondence: \texttt{asai@cbrc.jp}}\\[0.5em]
  \begin{minipage}{0.85\textwidth}\centering\small
    $^{1}$AI-Empowered Life Science Initiative (ALIS), Research Organization of
    Information and Systems, 4-1-1 Toranomon, Tokyo 105-6929, Japan\\[0.25em]
    $^{2}$School of Life Science and Technology, Institute of Science Tokyo,
    Tokyo, Japan
  \end{minipage}%
}
\date{}

\begin{document}
\maketitle
\begin{spacing}{1.0}

\textbf{Background:}
Forward--backward and inside--outside algorithms are usually introduced
separately for hidden Markov models, sequence alignment, stochastic
context-free grammars, and RNA secondary-structure ensembles. Their
backward and outside quantities nevertheless share a common interpretation
as adjoints of forward or inside variables. This differential view unifies
posterior inference and expected feature counts, but does not by itself
determine the exact effect of a finite sequence substitution.

\textbf{Methods:}
We formulate these models as sum--product dynamic programs and distinguish
posterior item use, posterior local events, expected feature counts,
parameter-learning derivatives, infinitesimal sensitivities, and finite
sequence replacements. Sequence substitutions are represented as finite
changes in sequence-dependent local factors. We characterize exact
replacement through multi-affinity in position-specific factor groups and
extend the analysis to context-dependent RNA thermodynamic models.

\textbf{Results:}
Reverse differentiation recovers the standard backward and outside
recursions and expresses posterior quantities and expected counts as
inside--outside contractions or logarithmic derivatives. Under
multi-affinity, the same adjoint-derived coefficients reconstruct finite
sequence substitutions exactly; multisite changes additionally require
mixed terms. Nearest-neighbor RNA thermodynamics generally falls outside
this single-factor-group regime because one nucleotide can alter
overlapping energy factors and boundary contexts. Exact RNA mutation
effects therefore require context-dependent inside--local--outside
recombination, as exemplified by Rchange. Numerical checks reproduce direct
partition-function recomputation to machine precision.

\textbf{Conclusions:}
Adjoint variables provide a common calculus for inference, expected-count
learning, sensitivity analysis, and finite sequence perturbation in
biological dynamic programming. Whether they determine an exact discrete
sequence effect is governed by the sequence-to-factor structure:
multi-affinity permits an exact finite expansion, whereas overlapping
context-dependent factors require broader recombination.


\end{spacing}

\section{Introduction}
\label{sec:introduction}

Dynamic programming is central to biological sequence analysis. Hidden
Markov models (HMMs), probabilistic sequence alignment, stochastic
context-free grammars (SCFGs), and RNA folding algorithms sum over
different latent objects, but their recursions share a common
sum--product form. A global probability or partition function is assembled
from local factors and partial forward or inside quantities, while
backward or outside quantities propagate information in the reverse
direction.

These reverse quantities have a well-established interpretation as
adjoints of the forward or inside variables: they are derivatives of the
global sum with respect to intermediate quantities in the dynamic program.
This viewpoint connects forward--backward and inside--outside algorithms to
semiring dynamic programming, sum--product message passing,
network-polynomial differentiation, reverse-mode automatic
differentiation, and structured prediction
\citep{Goodman1999,Kschischang2001,Darwiche2003,
GriewankWalther2008,Baydin2018,Eisner2016}. It also explains why the same
reverse computation yields posterior item use, posterior local-event
contributions, and expected feature counts. Expected-count derivatives can
in turn support parameter learning; coupled RNA grammar models such as
RNAelem provide one concrete example \citep{MiyakeEtAl2024}.

A different question arises when the observed biological sequence itself
is changed. Parameter differentiation varies model factors while the
sequence is fixed. Mutational analysis and sequence design instead replace
one or more observed symbols by finite amounts while the model is held
fixed. A derivative describes an infinitesimal perturbation, but a
nucleotide or amino-acid substitution changes a collection of
sequence-dependent factors at once. Several substitutions may also
interact within the same latent object. Consequently, differentiability of
a dynamic program does not by itself imply that a gradient gives the exact
endpoint effect of a discrete sequence edit.

The central question of this paper is therefore: under what conditions do
adjoint-derived coefficients determine an exact finite sequence
replacement? We address this question through the sequence-to-factor map
of the dynamic program. For each edited position, we collect the local
factor coordinates changed by the substitution and study the dependence of
the partition function on these position-specific groups. Multi-affinity
provides an exact finite expansion in first and mixed derivatives, whereas
failure of this factorization indicates that a broader recombination is
needed. The distinction concerns the partition function itself; nonlinear
quantities such as log-partition functions and posterior ratios require
separate treatment even when the underlying change in \(Z\) is recovered
exactly.

We examine this criterion in representative HMM, SCFG, alignment, and RNA
models. The RNA case is particularly informative because
nearest-neighbor thermodynamic factors overlap and depend on nucleotide
contexts at loop and interval boundaries. These dependencies motivate a
context-dependent inside--local--outside construction rather than a single
position-group contraction. We relate this construction to Rchange, which
computes mutation-induced partition-function changes by reusing
context-indexed RNA inside and outside quantities
\citep{Kiryu2012}. This relation places ordinary derivative
contractions and RNA mutation recombination within one finite-replacement
perspective without treating them as identical algorithms.

This paper makes three main contributions. First, it develops a common
adjoint calculus that distinguishes posterior items, local events,
expected feature counts, parameter-learning derivatives, and finite
sequence effects. Second, it establishes a model-independent
multi-affinity criterion for exact finite replacement, including the mixed
terms required for multiple substitutions. Third, it extends the
sequence-to-factor analysis to context-dependent RNA thermodynamics and
connects the resulting recombination principle to Rchange. Representative
numerical checks are used to confirm the derived identities and to
illustrate the difference between exact finite changes and common
first-order approximations.

Section~2 develops the adjoint and finite-replacement framework through
HMMs and SCFGs. Section~3 extends it to affine-gap alignment and RNA
ensembles, including expected-count learning in RNAelem and a
sequence-to-factor classification. Section~4 treats context-dependent RNA
mutation recombination and multiple mutations. Section~5 discusses
computational implications, numerical verification, limitations, and
broader connections. Detailed derivations, model-specific recursions,
RNA recombination formulas, multisite expansions, and reproducibility
information are provided in the Supplementary Material.

\section{A General Differential Framework through HMMs and SCFGs}
\label{sec:general-calculus}

We use hidden Markov models (HMMs) and stochastic context-free
grammars (SCFGs) to develop the common adjoint and finite-replacement
calculus. Complete derivations and technical details are given in
Supplementary Section~S1.

\subsection{HMMs, SCFGs, and a common sum--product form}
\label{sec:hmm-scfg-sum-product}

Let \(x=x_1\cdots x_L\) be an observed biological sequence. We first
specify the latent objects and global sums for the two models. For an HMM
with hidden-state set \(\mathcal{S}\), initial probabilities \(\pi_s\),
transition probabilities \(a_{st}\), and emission probabilities \(e_s(c)\),
the latent object is a state path \(z=(z_1,\ldots,z_L)\), with weight
\begin{equation}
  w_{\mathrm{HMM}}(z;x)
  =
  \pi_{z_1}e_{z_1}(x_1)
  \prod_{i=2}^{L}
  a_{z_{i-1}z_i}e_{z_i}(x_i),
  \label{eq:hmm-path-weight}
\end{equation}
and
\begin{equation}
  Z_{\mathrm{HMM}}(x)
  =
  \sum_{z\in\mathcal{S}^{L}}
  w_{\mathrm{HMM}}(z;x).
  \label{eq:hmm-sequence-probability}
\end{equation}
For an SCFG, let \(\mathcal{N}\) be the nonterminal set and \(S\) the start
symbol. We consider a grammar in Chomsky normal form for nonempty strings,
with no \(\epsilon\)-productions: each rule is either \(A\to BC\), with
weight \(t_{A\to BC}\), or \(A\to c\), with factor \(e_A(c)\). Let
\(\mathcal{N}_{\mathrm{lex}}\subseteq\mathcal{N}\) denote the nonterminals
with lexical rules. The latent objects are derivation trees
\(\tau\in\mathcal{T}(x)\) with yield \(x\), and
\begin{equation}
  Z_{\mathrm{SCFG}}(x)
  =
  \sum_{\tau\in\mathcal{T}(x)}
  w_{\mathrm{SCFG}}(\tau;x),
  \label{eq:scfg-string-probability}
\end{equation}
where \(w_{\mathrm{SCFG}}(\tau;x)\) is the product of the rule factors in
\(\tau\). Distinct derivation trees are treated as distinct latent objects,
even when the grammar is ambiguous.

The corresponding forward and inside recursions sum partial latent
objects. For the HMM,
\begin{equation}
  \alpha_i(t)
  =
  e_t(x_i)
  \sum_{s\in\mathcal{S}}
  \alpha_{i-1}(s)a_{st},
  \qquad i=2,\ldots,L,
  \label{eq:hmm-forward-recursion}
\end{equation}
with the usual initialization and
\(Z_{\mathrm{HMM}}(x)=\sum_s\alpha_L(s)\). For the SCFG, let
\(I_A(i,j)\) be the inside value for nonterminal \(A\) spanning
\(x_i\cdots x_j\). With \(I_A(i,i)=e_A(x_i)\), the binary recursion is
\begin{equation}
  I_A(i,j)
  =
  \sum_{A\to BC}
  t_{A\to BC}
  \sum_{k=i}^{j-1}
  I_B(i,k)I_C(k+1,j),
  \qquad i<j,
  \label{eq:scfg-inside-recursion}
\end{equation}
and \(Z_{\mathrm{SCFG}}(x)=I_S(1,L)\).

Both models are instances of the same sum--product construction. Let
\(\Omega\) be a finite set of latent objects, and let \(N_r(\omega)\) count
the occurrences of local decomposition \(r\) in \(\omega\). For
nonnegative local factors \(\phi_r\),
\begin{equation}
  Z
  =
  \sum_{\omega\in\Omega}
  w(\omega),
  \qquad
  w(\omega)
  =
  \prod_r
  \phi_r^{N_r(\omega)}.
  \label{eq:general-latent-sum}
\end{equation}
For each dynamic-programming item \(u\), let \(\mathcal{R}(u)\) be the
finite set of local decompositions contributing to \(u\). A decomposition
\(r\in\mathcal{R}(u)\) has local factor \(\phi_r\) and a finite list of
child items \(\operatorname{ch}(r)\). Repeated child occurrences retain
their multiplicity; for a terminal decomposition, this list is empty and
the empty product is one. A general acyclic sum--product recursion is
therefore
\begin{equation}
  A_u
  =
  \sum_{r\in\mathcal{R}(u)}
  \phi_r
  \prod_{v\in\operatorname{ch}(r)}
  A_v,
  \label{eq:general-inside}
\end{equation}
with the root item equal to \(Z\). HMM items are forward states
\((i,s)\), whereas SCFG items are nonterminal spans \((A,i,j)\).

\subsection{Backward and outside quantities as adjoints}
\label{sec:hmm-scfg-adjoints}

The HMM backward and SCFG outside quantities are derivatives of the global
sum with respect to forward or inside items:
\begin{equation}
  \beta_i(s)
  =
  \frac{\partial Z_{\mathrm{HMM}}}{\partial\alpha_i(s)},
  \qquad
  O_A(i,j)
  =
  \frac{\partial Z_{\mathrm{SCFG}}}{\partial I_A(i,j)}.
  \label{eq:hmm-scfg-adjoint-definitions}
\end{equation}
Reverse differentiation of Eq.~\eqref{eq:hmm-forward-recursion} gives the
standard HMM backward recursion
\begin{equation}
  \beta_i(s)
  =
  \sum_{t\in\mathcal{S}}
  a_{st}e_t(x_{i+1})\beta_{i+1}(t),
  \qquad
  i=1,\ldots,L-1.
  \label{eq:hmm-backward-recursion}
\end{equation}

For the SCFG considered here, an inside item \(I_A(i,j)\) can occur as
the left or right child of a binary parent production. For a non-root span,
reverse differentiation of Eq.~\eqref{eq:scfg-inside-recursion} gives
\begin{align}
  O_A(i,j)
  &=
  \sum_{B,C\in\mathcal{N}}
  \sum_{\ell=j+1}^{L}
  O_B(i,\ell)
  t_{B\to AC}
  I_C(j+1,\ell)
  \nonumber\\
  &\quad+
  \sum_{B,C\in\mathcal{N}}
  \sum_{h=1}^{i-1}
  O_B(h,j)
  t_{B\to CA}
  I_C(h,i-1).
  \label{eq:scfg-outside-recursion}
\end{align}
The first term collects parent decompositions in which \(A\) is the left
child, and the second those in which \(A\) is the right child. Standard
terminal and root boundary conditions, together with equivalent
reverse-accumulation update forms, are given in Supplementary
Sections~S1.2 and~S1.3.

In the general computation, define
\begin{equation}
  B_u
  =
  \frac{\partial Z}{\partial A_u}.
  \label{eq:general-adjoint}
\end{equation}
Reverse application of the multivariable chain rule gives
\begin{equation}
  B_u
  =
  \mathbf{1}\{u=u_{\mathrm{root}}\}
  +
  \sum_{p:\,A_p\text{ depends on }A_u}
  B_p
  \frac{\partial A_p}{\partial A_u}.
  \label{eq:general-reverse}
\end{equation}
Thus backward and outside algorithms are two instances of the same reverse
accumulation on a sum--product dynamic program.

\subsection{Posterior quantities and expected feature counts}
\label{sec:hmm-scfg-posterior-counts}

For a fixed observed sequence \(x\), the conditional distribution over
latent objects in an HMM or SCFG has an exponential-family form. More
generally, suppose that the positive factor parameters are
\(\{\psi_k\}\), and let \(N_k(\omega)\) count the total multiplicity of
factor \(\psi_k\) in latent object \(\omega\). With
\(\theta_k=\log\psi_k\),
\begin{equation}
  P(\omega\mid x;\boldsymbol{\theta})
  =
  \frac{w(\omega)}{Z(\boldsymbol{\theta})}
  =
  \exp\!\left\{
    \sum_k \theta_k N_k(\omega)
    -
    \log Z(\boldsymbol{\theta})
  \right\}.
  \label{eq:general-exponential-family}
\end{equation}
Thus the factor counts are sufficient statistics, and the standard
log-partition identity gives
\begin{equation}
  \frac{\partial\log Z}{\partial\theta_k}
  =
  \frac{\partial\log Z}{\partial\log\psi_k}
  =
  \mathbb{E}\!\left[N_k\mid x\right].
  \label{eq:general-expected-count}
\end{equation}
This identity underlies posterior item use, posterior local-event use, and
expected counts of shared model features
\citep{Darwiche2003,Eisner2016,WainwrightJordan2008}.

To obtain the expected use of a dynamic-programming item \(u\), introduce
an auxiliary marker \(\lambda_u\) that multiplies every occurrence of
\(A_u\) in the parent decompositions, and write
\(\eta_u=\log\lambda_u\). In the marked latent-object expansion,
\(\eta_u\) is the natural parameter associated with the sufficient
statistic \(N_u(\omega)\). Therefore,
\begin{equation}
  \mathbb{E}\!\left[N_u\mid x\right]
  =
  \left.
  \frac{\partial\log Z}{\partial\eta_u}
  \right|_{\eta_u=0}
  =
  \left.
  \frac{\partial\log Z}{\partial\log\lambda_u}
  \right|_{\lambda_u=1}
  =
  \frac{A_u}{Z}
  \frac{\partial Z}{\partial A_u}
  =
  \frac{A_uB_u}{Z}.
  \label{eq:general-item-use}
\end{equation}
The last equality uses the adjoint definition
\(B_u=\partial Z/\partial A_u\). Thus the familiar
inside--outside or forward--backward product is an exponential-family
expected-count identity expressed through the dynamic-programming
computational graph.

A particular local decomposition \(r\in\mathcal{R}(u)\) has forward
contribution
\begin{equation}
  T_r
  =
  \phi_r
  \prod_{v\in\operatorname{ch}(r)}
  A_v.
  \label{eq:general-local-term}
\end{equation}
Introducing a marker \(\lambda_r\) that replaces \(T_r\) by
\(\lambda_rT_r\) gives, by the same argument,
\begin{equation}
  \mathbb{E}\!\left[N_r\mid x\right]
  =
  \left.
  \frac{\partial\log Z}{\partial\log\lambda_r}
  \right|_{\lambda_r=1}
  =
  \frac{B_uT_r}{Z}.
  \label{eq:general-local-event-use}
\end{equation}
Equation~\eqref{eq:general-item-use} concerns use of an item, whereas
Eq.~\eqref{eq:general-local-event-use} selects one particular local
decomposition creating that item. If a positive model factor is shared by
several decompositions, Eq.~\eqref{eq:general-expected-count} sums the
corresponding local-event contributions and gives its total expected
feature count.

The HMM and SCFG formulas are immediate special cases. For an HMM state
item \((i,s)\), \(N_{(i,s)}(z)\) is the indicator of \(z_i=s\), and
Eq.~\eqref{eq:general-item-use} gives
\begin{equation}
  P(z_i=s\mid x)
  =
  \frac{\alpha_i(s)\beta_i(s)}{Z_{\mathrm{HMM}}}.
  \label{eq:hmm-state-posterior}
\end{equation}
For the local transition--emission event
\(z_{i-1}=s,z_i=t\), Eq.~\eqref{eq:general-local-event-use} gives
\begin{equation}
  P(z_{i-1}=s,z_i=t\mid x)
  =
  \frac{
    \alpha_{i-1}(s)a_{st}e_t(x_i)\beta_i(t)
  }{
    Z_{\mathrm{HMM}}
  }.
  \label{eq:hmm-transition-posterior}
\end{equation}

For an SCFG item \((A,i,j)\), the item count indicates whether the
derivation contains a constituent \(A\) spanning \(x_i\cdots x_j\), so
\begin{equation}
  P\!\left(
    A\Rightarrow^{\ast}x_i\cdots x_j
    \mid x
  \right)
  =
  \frac{I_A(i,j)O_A(i,j)}{Z_{\mathrm{SCFG}}}.
  \label{eq:scfg-constituent-posterior}
\end{equation}
For the local event using \(A\to BC\) at split point \(k\) on span
\((i,j)\),
\begin{equation}
P\!\left(
A \to BC \text{ at } (i,k,j)
\mid x
\right)
=
\frac{
O_A(i,j)t_{A\to BC}I_B(i,k)I_C(k+1,j)
}{
Z_{\mathrm{SCFG}}
}.
\label{eq:scfg-local-posterior}
\end{equation}
Formal proofs and the treatment of repeated item occurrences, shared
parameters, and zero-valued factors are given in Supplementary
Section~S1.4.

\subsection{Finite sequence substitutions are not generally first derivatives}
\label{sec:finite-substitution-not-general-derivative}

Suppose that a substitution at position \(p\) replaces a vector of
sequence-dependent local factors \(\boldsymbol{\rho}_p\) by
\(\boldsymbol{\rho}'_p=\boldsymbol{\rho}_p+
\Delta\boldsymbol{\rho}_p\). The finite partition-function change is
\begin{equation}
  \Delta_p Z
  =
  Z\!\left(\boldsymbol{\rho}'_p\right)
  -
  Z\!\left(\boldsymbol{\rho}_p\right),
  \label{eq:general-finite-one-site-change}
\end{equation}
with all other factors held fixed. Derivatives evaluated at the original
factors give the first-order contraction
\begin{equation}
  \sum_{a}
  \frac{\partial Z}{\partial\rho_{p,a}}
  \Delta\rho_{p,a}.
  \label{eq:general-first-order-factor-contraction}
\end{equation}
Differentiability alone does not make
Eqs.~\eqref{eq:general-finite-one-site-change} and
\eqref{eq:general-first-order-factor-contraction} equal. For example, if
\begin{equation}
  Z(\rho)=\rho^2,
\end{equation}
then
\begin{equation}
  Z(\rho+\Delta\rho)-Z(\rho)
  =
  2\rho\,\Delta\rho
  +
  (\Delta\rho)^2,
  \label{eq:general-quadratic-counterexample}
\end{equation}
whereas the original-point derivative contributes only
\(2\rho\,\Delta\rho\). In a dynamic program, the missing nonlinear terms
can arise when several factors controlled by the same sequence position are
used jointly in one latent object. Thus a finite sequence edit is not, in
general, reconstructed exactly by a first derivative at the original
sequence.

HMMs and the SCFG considered here have an additional exactly-once property
that removes this nonlinearity for a one-position substitution. They
therefore provide the simplest exact cases and motivate the general
factor-group criterion in Section~\ref{sec:general-multi-affinity}.

\subsection{Exact one-position substitutions in HMMs and SCFGs}
\label{sec:hmm-scfg-one-site-substitutions}

For the HMM and SCFG, define the position-specific factor groups
\begin{equation}
  \boldsymbol{\rho}^{\mathrm{HMM}}_i
  =
  \left(e_s(x_i)\right)_{s\in\mathcal{S}},
  \qquad
  \boldsymbol{\rho}^{\mathrm{SCFG}}_i
  =
  \left(e_A(x_i)\right)_{A\in\mathcal{N}_{\mathrm{lex}}}.
  \label{eq:hmm-scfg-position-factor-groups}
\end{equation}
Every HMM path emits position \(i\) in exactly one hidden state. Hence,
with all other factors fixed, its partition function can be written as
\begin{equation}
  Z_{\mathrm{HMM}}
  =
  \sum_{s\in\mathcal{S}}
  C^{\mathrm{HMM}}_{i,s}
  \rho^{\mathrm{HMM}}_{i,s},
  \label{eq:hmm-position-linear-decomposition}
\end{equation}
where the coefficients \(C^{\mathrm{HMM}}_{i,s}\) are independent of the
entire group \(\boldsymbol{\rho}^{\mathrm{HMM}}_i\). Likewise, every
complete derivation of the binary--lexical SCFG considered here emits
position \(i\) through exactly one lexical event, so
\begin{equation}
  Z_{\mathrm{SCFG}}
  =
  \sum_{A\in\mathcal{N}_{\mathrm{lex}}}
  C^{\mathrm{SCFG}}_{i,A}
  \rho^{\mathrm{SCFG}}_{i,A},
  \label{eq:scfg-position-linear-decomposition}
\end{equation}
with coefficients independent of
\(\boldsymbol{\rho}^{\mathrm{SCFG}}_i\).

The coefficients in
Eqs.~\eqref{eq:hmm-position-linear-decomposition} and
\eqref{eq:scfg-position-linear-decomposition} are the corresponding
partial derivatives. Therefore, if \(x^{i\to c}\) is obtained by replacing
\(x_i\) by \(c\),
\begin{align}
  Z_{\mathrm{HMM}}\!\left(x^{i\to c}\right)
  -
  Z_{\mathrm{HMM}}(x)
  &={}
  \sum_{s\in\mathcal{S}}
  \frac{\partial Z_{\mathrm{HMM}}}
       {\partial\rho^{\mathrm{HMM}}_{i,s}}
  \left[e_s(c)-e_s(x_i)\right],
  \label{eq:hmm-one-site-substitution}
  \\
  Z_{\mathrm{SCFG}}\!\left(x^{i\to c}\right)
  -
  Z_{\mathrm{SCFG}}(x)
  &={}
  \sum_{A\in\mathcal{N}_{\mathrm{lex}}}
  \frac{\partial Z_{\mathrm{SCFG}}}
       {\partial\rho^{\mathrm{SCFG}}_{i,A}}
  \left[e_A(c)-e_A(x_i)\right].
  \label{eq:scfg-one-site-substitution}
\end{align}
The derivative coefficients are obtained from the same backward or outside
calculation used for posterior inference. In particular,
\(\partial Z_{\mathrm{SCFG}}/
\partial\rho^{\mathrm{SCFG}}_{i,A}=O_A(i,i)\). The complete HMM
coefficient and both linear decompositions are derived in Supplementary
Section~S1.5.

These formulas are exact because the partition functions are linear in the
factor group changed at position \(i\), not because the substitutions are
small. This observation suggests that exactness is controlled by the
algebraic dependence of \(Z\) on the mutation-dependent factor groups,
rather than by whether the latent objects are paths or trees.

\subsection{Group affinity and a general multi-affinity criterion}
\label{sec:general-multi-affinity}

For each affected position \(p\), let
\begin{equation}
  \boldsymbol{\rho}_p
  =
  \left(\rho_{p,a}\right)_{a\in\mathcal{A}_p}
  \label{eq:general-factor-group}
\end{equation}
collect all sequence-dependent factor coordinates changed by the
substitution. We say that \(Z\) is \emph{affine in the factor group}
\(\boldsymbol{\rho}_p\) if, with all other factor groups fixed,
\begin{equation}
  Z
  =
  C_p
  +
  \sum_{a\in\mathcal{A}_p}
  C_{p,a}\rho_{p,a},
  \label{eq:general-group-affine}
\end{equation}
where \(C_p\) and \(C_{p,a}\) are independent of
\(\boldsymbol{\rho}_p\). Group-wise affinity is sufficient for an exact
one-position replacement:
\begin{equation}
  \Delta_p Z
  =
  \sum_{a\in\mathcal{A}_p}
  \frac{\partial Z}{\partial\rho_{p,a}}
  \Delta\rho_{p,a}.
  \label{eq:general-one-site-replacement}
\end{equation}
The constant term \(C_p\) does not affect the difference. Thus homogeneity
is not required for one-site exactness.

The dependence is \emph{homogeneous of degree one in the group}
\(\boldsymbol{\rho}_p\) if \(C_p=0\). Equivalently, with all other
variables fixed,
\begin{equation}
  Z\!\left(
    \ldots,
    \lambda\boldsymbol{\rho}_p,
    \ldots
  \right)
  =
  \lambda
  Z\!\left(
    \ldots,
    \boldsymbol{\rho}_p,
    \ldots
  \right)
  \label{eq:general-group-homogeneity}
\end{equation}
for every scalar \(\lambda\). The exactly-once property of HMMs and the
SCFG considered above gives this stronger homogeneous form. Other models
can be affine but non-homogeneous.

Finally, \(Z\) is \emph{multi-affine} in a collection of factor groups if
it is affine in each group while the remaining groups are held fixed. Let
\(\mathcal{P}\) be the set of substituted positions and
\(\boldsymbol{\rho}'_p=\boldsymbol{\rho}_p+
\Delta\boldsymbol{\rho}_p\). For \(U\subseteq\mathcal{P}\), define
\(\mathcal{A}_U=\prod_{p\in U}\mathcal{A}_p\).

\begin{proposition}[Exact finite replacement under multi-affinity]
\label{prop:multi-affine-replacement}
Suppose that \(Z\) is multi-affine in
\(\{\boldsymbol{\rho}_p:p\in\mathcal{P}\}\). Then
\begin{align}
  Z(\boldsymbol{\rho}')-Z(\boldsymbol{\rho})
  &={}
  \sum_{\emptyset\neq U\subseteq\mathcal{P}}
  \;
  \sum_{\boldsymbol{a}_U\in\mathcal{A}_U}
  \frac{
    \partial^{\lvert U\rvert}Z
  }{
    \displaystyle
    \prod_{p\in U}\partial\rho_{p,a_p}
  }
  \prod_{p\in U}
  \Delta\rho_{p,a_p},
  \label{eq:general-multi-affine-expansion}
\end{align}
where all derivatives are evaluated at the original factors.
\end{proposition}

The result follows by expanding successively in the affected affine groups;
a complete proof is given in Supplementary Section~S1.7. For one position,
the proposition reduces to Eq.~\eqref{eq:general-one-site-replacement}. For
two distinct positions \(p<q\),
\begin{align}
  \Delta_{p,q}Z
  &={}
  \Delta_p Z
  +
  \Delta_q Z
  +
  \sum_{a\in\mathcal{A}_p}
  \sum_{b\in\mathcal{A}_q}
  \frac{
    \partial^2 Z
  }{
    \partial\rho_{p,a}\,\partial\rho_{q,b}
  }
  \Delta\rho_{p,a}
  \Delta\rho_{q,b}.
  \label{eq:general-two-site-replacement}
\end{align}
Equivalently, the mixed finite interaction is
\begin{equation}
  \Delta_p\Delta_q Z
  =
  Z_{pq}-Z_p-Z_q+Z.
  \label{eq:general-mixed-finite-difference}
\end{equation}
Higher-order substitutions require the corresponding mixed derivatives for
all nonempty subsets of affected positions. Multi-affinity is a general
sufficient criterion for this exact finite expansion; it is not asserted to
be necessary for every possible representation or special cancellation.

\subsection{Scope of the criterion and transition to alignment and RNA}
\label{sec:scope-and-transition}

The exactness result concerns \(Z\), not nonlinear transformations of
\(Z\). Even when Eq.~\eqref{eq:general-one-site-replacement} is exact,
\begin{equation}
  \Delta_p\log Z
  =
  \log\!\left(
    1+
    \frac{\Delta_p Z}{Z}
  \right)
  \label{eq:general-log-change}
\end{equation}
remains nonlinear. Posterior ratios and other normalized quantities
likewise need not have exact first-order replacement formulas.

The criterion depends on the sequence-to-factor map, not only on the
dynamic-programming topology. In alignment models, the partition function
can remain affine in a mutation-dependent factor group without being
homogeneous in that group. The precise form depends on the alignment
parameterization and is discussed in Section~3.

In a nearest-neighbor RNA energy model, one nucleotide can change several
factors that occur jointly and can alter mutation-sensitive boundary
contexts. The adjoint interpretation remains valid, but the edit is not
generally a single affine position-group replacement. Exact RNA mutation
effects therefore require context-dependent finite recombination rather
than one contraction of original-background derivatives.

Section~3 develops the alignment and RNA partition-function adjoints, and
Section~4 treats context-dependent RNA mutation recombination.

\section{Extensions to Alignment and RNA Ensemble Models}
\label{sec:extensions-alignment-rna}

This section extends the framework of Section~2 to alignment and RNA
ensemble models, and illustrates its connection to RNA
sequence--structure learning.

\subsection{Affine-gap alignment ensembles}
\label{sec:alignment}

Alignment provides a useful intermediate case between the exactly-once
models of Section~2 and the context-dependent RNA models considered below.
The relevant point is not the lattice recursion itself, but how a sequence
substitution enters the alignment partition function.

Let \(x=x_1\cdots x_L\) and \(y=y_1\cdots y_K\). For a matched pair, use
the positive factor
\begin{equation}
  q(a,b)
  =
  \exp\!\left\{
    \frac{s(a,b)}{\tau}
  \right\},
  \label{eq:alignment-match-factor}
\end{equation}
with the usual affine-gap factors
\(p_d=\exp(-d/\tau)\) and \(p_e=\exp(-e/\tau)\).
Summing the weights of all global alignment paths gives the partition
function \(Z_{\mathrm{aln}}(x,y)\). The standard three-state affine-gap
recursions are sum--product dynamic programs, so their reverse variables are
again adjoints of the forward variables. In particular, if \(F_M(i,j)\) is
the total prefix weight of paths whose latest event matches \(x_i\) with
\(y_j\), define
\begin{equation}
  B_M(i,j)
  =
  \frac{
    \partial Z_{\mathrm{aln}}(x,y)
  }{
    \partial F_M(i,j)
  }.
  \label{eq:alignment-match-adjoint}
\end{equation}
The corresponding match posterior is
\begin{equation}
  P(x_i\sim y_j\mid x,y)
  =
  \frac{
    F_M(i,j)B_M(i,j)
  }{
    Z_{\mathrm{aln}}(x,y)
  }.
  \label{eq:alignment-match-posterior}
\end{equation}
Thus alignment uses the same adjoint and expected-count calculus as the
models in Section~2. Complete forward and reverse recursions, gap-event
posteriors, and score derivatives are given in Supplementary Section~S2.

Now consider a substitution \(x_i\to c\). In the score-based model above,
the sequence-identity-dependent factors changed at position \(i\) are
\begin{equation}
  \boldsymbol{\rho}^{M}_i
  =
  \left(
    \rho^{M}_{i,j}
  \right)_{j=1}^{K},
  \qquad
  \rho^{M}_{i,j}
  =
  q(x_i,y_j).
  \label{eq:alignment-position-group}
\end{equation}
Every alignment path uses at most one coordinate from this group: it either
matches \(x_i\) to one \(y_j\), or aligns \(x_i\) to a gap. Consequently,
with all other factors fixed,
\begin{equation}
  Z_{\mathrm{aln}}
  =
  C_i
  +
  \sum_{j=1}^{K}
  C_{i,j}\rho^{M}_{i,j},
  \label{eq:alignment-affine-form}
\end{equation}
where \(C_i\) and \(C_{i,j}\) are independent of the whole group
\(\boldsymbol{\rho}^{M}_i\). The partition function is therefore affine in
the mutation-dependent factor group. Moreover,
\begin{equation}
  C_{i,j}
  =
  \frac{
    \partial Z_{\mathrm{aln}}
  }{
    \partial\rho^{M}_{i,j}
  }
  =
  \frac{
    F_M(i,j)B_M(i,j)
  }{
    q(x_i,y_j)
  }.
  \label{eq:alignment-affine-coefficient}
\end{equation}

The one-site finite substitution is therefore exact:
\begin{align}
  &
  Z_{\mathrm{aln}}\!\left(
    x^{i\to c},y
  \right)
  -
  Z_{\mathrm{aln}}(x,y)
  \nonumber\\
  &\qquad=
  \sum_{j=1}^{K}
  \frac{
    F_M(i,j)B_M(i,j)
  }{
    q(x_i,y_j)
  }
  \left[
    q(c,y_j)-q(x_i,y_j)
  \right].
  \label{eq:alignment-exact-substitution}
\end{align}
Equivalently,
\begin{equation}
  \frac{
    Z_{\mathrm{aln}}\!\left(
      x^{i\to c},y
    \right)
    -
    Z_{\mathrm{aln}}(x,y)
  }{
    Z_{\mathrm{aln}}(x,y)
  }
  =
  \sum_{j=1}^{K}
  P(x_i\sim y_j\mid x,y)
  \left[
    \frac{q(c,y_j)}{q(x_i,y_j)}
    -
    1
  \right].
  \label{eq:alignment-exact-substitution-posterior}
\end{equation}
The equality is exact because of group affinity, not because the
substitution is small.

Alignment also shows why affinity and homogeneity should be distinguished.
In the factor group of Eq.~\eqref{eq:alignment-position-group}, paths that
align \(x_i\) to a gap use no coordinate from the group, so the constant
term \(C_i\) in Eq.~\eqref{eq:alignment-affine-form} need not vanish.
The dependence is therefore affine but not necessarily homogeneous. This
is parameterization-dependent: in a pair-HMM with residue-dependent
gap-emission factors, the complete mutation-dependent group must include
those factors as well, and an expanded group can recover an exactly-once,
degree-one homogeneous representation. The invariant statement is that an
exact one-site substitution follows whenever the complete set of factors
changed by the edit forms an affine group. Parameterization variants and
multisite expansions are detailed in Supplementary Section~S2.

\subsection{RNA secondary-structure ensembles}
\label{sec:rna-ensemble}

RNA secondary-structure ensembles obey the same adjoint calculus as the
models above, but they differ sharply in how a sequence substitution enters
the local factors. Let \(x=x_1\cdots x_L\), and let \(\Omega_L\) be a common
set of pseudoknot-free secondary-structure candidates on the labelled
positions. For a local structural configuration \(\ell\), define
\begin{equation}
  \phi_x(\ell)
  =
  \begin{cases}
    \exp\!\left\{
      -E_x(\ell)/(RT)
    \right\},
    &
    \ell\text{ is compatible with }x,
    \\[4pt]
    0,
    &
    \text{otherwise}.
  \end{cases}
  \label{eq:rna-local-factor}
\end{equation}
The zero-factor convention keeps the latent-structure universe fixed across
sequence substitutions. The partition function is
\begin{equation}
  Z_{\mathrm{RNA}}(x)
  =
  \sum_{\sigma\in\Omega_L}
  \prod_{\ell\in\sigma}
  \phi_x(\ell).
  \label{eq:rna-partition}
\end{equation}

McCaskill-type recursions evaluate this sum through interval quantities
\citep{McCaskill1990}. Let \(Z^b(i,j)\) denote the inside partition
function conditional on base pair \((i,j)\), and define its outside adjoint
by
\begin{equation}
  W^b(i,j)
  =
  \frac{
    \partial Z_{\mathrm{RNA}}(x)
  }{
    \partial Z^b(i,j)
  }.
  \label{eq:rna-paired-outside}
\end{equation}
The standard base-pairing probability is therefore
\begin{equation}
  P(i,j\mid x)
  =
  \frac{
    Z^b(i,j)W^b(i,j)
  }{
    Z_{\mathrm{RNA}}(x)
  }.
  \label{eq:rna-base-pair-posterior}
\end{equation}
Thus the familiar RNA inside--outside product is the same normalized
item-adjoint identity as Eq.~\eqref{eq:general-item-use}. Likewise, if the
fixed-sequence energy is written as
\begin{equation}
  E_x(\sigma)
  =
  \sum_k
  \theta_k f_k(\sigma,x),
  \qquad
  \eta_k
  =
  -\frac{\theta_k}{RT},
  \label{eq:rna-energy-feature-form}
\end{equation}
then
\begin{equation}
  \frac{
    \partial\log Z_{\mathrm{RNA}}
  }{
    \partial\eta_k
  }
  =
  \mathbb{E}\!\left[
    f_k
    \mid x
  \right].
  \label{eq:rna-expected-feature}
\end{equation}
Detailed decomposition-level recursions, outside updates, local-event
posteriors, and parameter sensitivities are given in Supplementary
Section~S3.

The distinction appears when the sequence itself is changed. In a
nearest-neighbor energy model, one nucleotide can alter several local
Boltzmann factors that occur jointly in the same structure, and it can also
change nucleotide contexts needed at loop or interval boundaries. The
mutation-dependent terms therefore do not generally form one affine
position-specific factor group. The outside quantities remain valid
adjoints, but an exact nucleotide substitution requires the
context-dependent finite recombination developed in Section~4.

\subsection{Coupled grammars and learnable sequence--structure motifs}
\label{sec:coupled-grammars}

Coupled grammars provide a general way to combine two structured
factorizations within one sum--product dynamic program. Rather than treating
two latent structures independently, a product-style construction pairs only
compatible states and local decompositions. Abstractly, if
\(\omega_1\in\Omega_1\) and \(\omega_2\in\Omega_2\) are latent objects of
two structured models, the coupled latent space can be written as
\begin{equation}
  \Omega_{\mathrm{cpl}}
  =
  \left\{
    (\omega_1,\omega_2)
    \in
    \Omega_1\times\Omega_2
    :
    \omega_1\sim\omega_2
  \right\},
  \label{eq:coupled-latent-space}
\end{equation}
where \(\sim\) denotes model-specific compatibility. A coupled object then
inherits local factors from both components, so its total weight remains a
product of local factors. Once the compatible product construction is itself
represented as a finite sum--product dynamic program, the adjoint and
expected-count identities of Section~2 apply without modification. This is
a product construction over compatible derivations or states, not a claim
that arbitrary context-free languages can be intersected while remaining
context-free.

RNAelem is one biological instance of this idea
\citep{MiyakeEtAl2024}. It couples a Profile CFG representing a
sequence--structure motif with a CFG representation of the RNA energy model.
Let \(\varphi\) and \(\sigma\) denote compatible parse trees of the two
component grammars and let
\(\tau=\sigma\otimes\varphi\) denote the corresponding Coupled-CFG parse
tree. For model parameters
\begin{equation}
  \nu
  =
  (\mu,\boldsymbol{\theta},\lambda),
  \label{eq:rnaelem-parameters}
\end{equation}
RNAelem assigns a coupled weight containing both the learnable profile
factors and the thermodynamic contribution. If \(Z_\nu(x)\) is the full
coupled partition function and \(Z_{\nu,y}(x)\) is the partition function
restricted by the observed motif label \(y\), then
\begin{equation}
  P(y\mid\nu,x)
  =
  \frac{
    Z_{\nu,y}(x)
  }{
    Z_\nu(x)
  }.
  \label{eq:rnaelem-conditional-probability}
\end{equation}

For a log-linear Profile-CFG parameter \(\theta_k\) with sufficient
statistic \(N_k\), differentiation gives
\begin{equation}
  \frac{
    \partial\log P(y\mid\nu,x)
  }{
    \partial\theta_k
  }
  =
  \mathbb{E}_{\nu,y}\!\left[
    N_k
    \mid x
  \right]
  -
  \mathbb{E}_{\nu}\!\left[
    N_k
    \mid x
  \right].
  \label{eq:rnaelem-expected-count-gradient}
\end{equation}
Thus learning reduces to a difference between constrained and
unconstrained expected counts, both obtained by inside--outside calculations
on the coupled model.

RNAelem therefore illustrates a second use of the framework beyond finite
sequence replacement: structured models can first be combined into a new
sum--product system, after which the same adjoint calculus supports
parameter learning. Here the observed sequence is fixed and the model
parameters vary. The finite-substitution problem studied in Sections~2.4--4
instead changes the sequence itself and therefore requires a separate
analysis of the sequence-to-factor map. Further details of the RNAelem
construction and its expected-count derivatives are given in Supplementary
Section~S3.

\subsection{Sequence-to-factor classification}
\label{sec:sequence-factor-classification}

The examples above show that exact finite sequence replacement is governed
not by whether the latent object is a path, derivation tree, or secondary
structure, but by how a sequence edit maps to the local factors of the
dynamic program. For an edited position, the central question is whether
the complete set of factors changed by the edit forms an affine group in
the partition function. Table~\ref{tab:sequence-factor-classification}
summarizes this sequence-to-factor classification for the four
finite-substitution models considered here.

\begin{table}[t]
  \centering
  \small
  \caption{
    Sequence-to-factor classification of the four models considered for
    finite sequence substitution. Alignment homogeneity depends on the
    complete mutation-dependent factor group. RNAelem is omitted because it
    represents fixed-sequence parameter learning rather than a distinct
    finite-substitution regime.
  }
  \label{tab:sequence-factor-classification}
  \begin{tabular}{
    >{\raggedright\arraybackslash}p{0.14\textwidth}
    >{\raggedright\arraybackslash}p{0.21\textwidth}
    >{\raggedright\arraybackslash}p{0.24\textwidth}
    >{\raggedright\arraybackslash}p{0.28\textwidth}
  }
    \hline
    Model
    &
    Use of a position group in one latent object
    &
    Algebraic structure
    &
    Exact one-site sequence effect
    \\
    \hline
    HMM
    &
    Exactly one emission coordinate
    &
    Homogeneous multi-affine
    &
    One contraction of adjoint-derived coefficients
    \\
    SCFG
    &
    Exactly one lexical coordinate
    &
    Homogeneous multi-affine
    &
    One contraction of outside-derived coefficients
    \\
    Alignment
    &
    At most one coordinate from the selected mutation-dependent group
    &
    Affine; not necessarily homogeneous
    &
    One contraction after all mutation-dependent factors are included
    \\
    Nearest-neighbor RNA
    &
    Several overlapping local factors and boundary contexts may change
    &
    Not generally one affine position group
    &
    Context-dependent inside--local--outside recombination
    \\
    \hline
  \end{tabular}
\end{table}

HMMs and the binary--lexical SCFG satisfy the stronger exactly-once
condition, whereas alignment requires only the at-most-once condition for
the selected mutation-dependent group. Nearest-neighbor RNA falls outside
this simple regime because one nucleotide can alter several jointly used
factors and mutation-sensitive boundary contexts.

RNAelem is not a fifth finite-substitution class. Section~3.3 illustrates
an orthogonal use of the same adjoint calculus: structured models are first
combined into a coupled sum--product system, and expected counts are then
used for parameter learning while the observed sequence is fixed.
Section~\ref{sec:rna-recombination} develops the context-dependent finite
recombination required for RNA sequence changes beyond the single-group
affine regime.

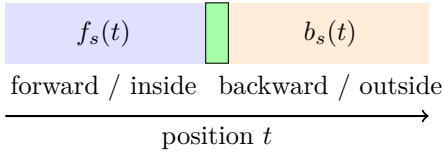
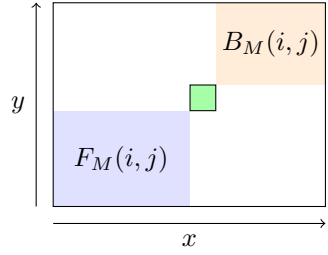
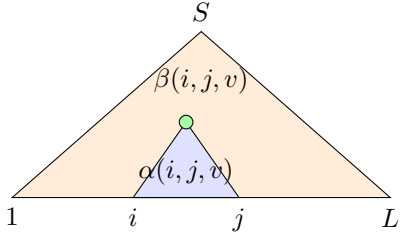
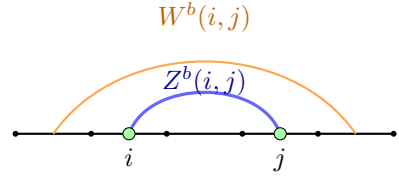
\begin{figure*}[t]
\centering
\tikzset{
  inside/.style ={fill=blue!12},
  outside/.style={fill=orange!15},
  selected/.style={draw, fill=green!35, inner sep=1.2pt},
  seq/.style    ={thick},
  font=\footnotesize,
}

\fbox{\parbox{0.94\textwidth}{\centering
$\displaystyle B_u=\frac{\partial Z}{\partial A_u}$\quad
$\displaystyle \frac{A_uB_u}{Z}=\mathbb E[N_u]$\quad
$\displaystyle \frac{B_u\phi_r\prod_{v\in\operatorname{ch}(r)}A_v}{Z}
=\mathbb E[N_r]$\\[1mm]
$\displaystyle \frac{\partial\log Z}{\partial\log\phi_k}
=\mathbb E[N_k]$
}}

\vspace{3mm}

\begin{minipage}[t]{0.48\textwidth}\centering
\begin{tikzpicture}
  \fill[inside]  (0,0) rectangle (2.65,0.8);
  \fill[outside] (2.95,0) rectangle (5.6,0.8);
  \node[selected,minimum height=0.8cm,minimum width=0.3cm] at (2.8,0.4){};
  \node at (1.3,0.4){$f_s(t)$};
  \node at (4.3,0.4){$b_s(t)$};
  \node[below] at (1.3,0){forward / inside};
  \node[below] at (4.3,0){backward / outside};
  \draw[seq,->] (0,-0.7)--(5.6,-0.7);
  \node[below] at (2.8,-0.7){position $t$};
\end{tikzpicture}\\[2pt]
(a) HMM: $P(\pi_t=s\mid x)=f_s(t)b_s(t)/P(x)$
\end{minipage}\hfill
\begin{minipage}[t]{0.48\textwidth}\centering
\begin{tikzpicture}[scale=0.9]
  \fill[inside]  (0,0) rectangle (2.0,1.4);
  \fill[outside] (2.4,1.8) rectangle (4.0,3.0);
  \draw (0,0) rectangle (4.0,3.0);
  \node[selected,minimum size=0.34cm] at (2.2,1.6){};
  \node at (1.0,0.7){$F_M(i,j)$};
  \node at (3.2,2.4){$B_M(i,j)$};
  \draw[->] (-0.25,0)--(-0.25,3.0) node[left,midway]{$y$};
  \draw[->] (0,-0.25)--(4.0,-0.25) node[below,midway]{$x$};
\end{tikzpicture}\\[2pt]
(b) Alignment: $P(x_i\sim y_j\mid x,y)=F_MB_M/Z(x,y)$
\end{minipage}

\vspace{4mm}

\begin{minipage}[t]{0.48\textwidth}\centering
\begin{tikzpicture}
  \fill[outside] (0,0)--(5,0)--(2.5,2.2)--cycle;
  \fill[inside]  (1.6,0)--(3.0,0)--(2.3,1.0)--cycle;
  \draw (0,0)--(5,0)--(2.5,2.2)--cycle;
  \draw (1.6,0)--(3.0,0)--(2.3,1.0)--cycle;
  \node[selected,circle,minimum size=0.18cm] at (2.3,1.0){};
  \node at (2.5,1.55){$\beta(i,j,v)$};
  \node at (2.3,0.35){$\alpha(i,j,v)$};
  \node[above] at (2.5,2.2){$S$};
  \node[below] at (0,0){$1$}; \node[below] at (5,0){$L$};
  \node[below] at (1.6,0){$i$}; \node[below] at (3.0,0){$j$};
\end{tikzpicture}\\[2pt]
(c) SCFG: $P([i,j],v\mid x)=\alpha(i,j,v)\beta(i,j,v)/P(x)$
\end{minipage}\hfill
\begin{minipage}[t]{0.48\textwidth}\centering
\begin{tikzpicture}
  \draw[seq] (0,0)--(5,0);
  \foreach \x in {0,1,2,3,4,5} \fill (\x,0) circle (1.1pt);
  \draw[orange!70,thick] (0.5,0) to[bend left=55] (4.5,0);
  \node[orange!75!black] at (2.5,1.55){$W^b(i,j)$};
  \draw[blue!60,very thick] (1.5,0) to[bend left=70] (3.5,0);
  \node[blue!60!black] at (2.5,0.7){$Z^b(i,j)$};
  \node[selected,circle,minimum size=0.16cm] at (1.5,0){};
  \node[selected,circle,minimum size=0.16cm] at (3.5,0){};
  \node[below] at (1.5,-0.05){$i$}; \node[below] at (3.5,-0.05){$j$};
\end{tikzpicture}\\[2pt]
(d) RNA: $P(i,j\mid x)=Z^b(i,j)W^b(i,j)/Z(x)$
\end{minipage}

\caption{Adjoint anatomy shared by the four models.  The panels show item
marginals: an inside or forward item $A_u$ is paired with its outside or
backward adjoint $B_u$, and $A_uB_u/Z$ gives the posterior expected use of
the selected item (green).  The boxed relations additionally show the
posterior of a particular local decomposition $r$, which contains its local
factor $\phi_r$ and child inside values.  Blue denotes inside/forward
contributions and orange denotes outside/backward contexts.}
\label{fig:unified}
\end{figure*}

\FloatBarrier
\section{Context-Dependent RNA Mutation Recombination}
\label{sec:rna-recombination}

Section~3.2 showed that RNA partition functions obey the same adjoint
calculus as the other models, while Section~3.4 identified
nearest-neighbor RNA as the case in which a nucleotide substitution does
not generally correspond to one affine position-specific factor group.
This section describes the broader finite-recombination principle needed
in that setting and relates it to Rchange \citep{Kiryu2012}.

\subsection{From local adjoints to context-dependent recombination}
\label{sec:rna-one-group-failure}

For a local RNA decomposition \(r\) creating item \(u\), the
fixed-sequence inside--local--outside contribution is
\begin{equation}
  T_r(x)
  =
  B_u(x)\,
  \phi_r(x)
  \prod_{v\in\operatorname{ch}(r)}
  A_v(x).
  \label{eq:rna-original-local-contribution}
\end{equation}
If a sequence edit changed only the exposed factor \(\phi_r\), the
corresponding finite effect would be obtained by replacing that factor
while reusing the same inside and outside quantities. Nearest-neighbor RNA
does not generally have this form. One nucleotide can alter several
primitive thermodynamic factors within the same decomposition, and it can
change the nucleotide contexts needed to evaluate adjacent inside or
outside states.

The exact construction must therefore expose the complete
mutation-dependent local block and evaluate the reusable pieces in contexts
consistent with the candidate nucleotide. Conceptually, each mutation case
has the form
\begin{equation}
  \text{outside context}
  \;\times\;
  \text{joint mutant local block}
  \;\times\;
  \text{compatible inside children}.
  \label{eq:rna-recombination-principle}
\end{equation}
The cases must partition the latent objects without overlap or omission.
This is the same inside--local--outside factorization used for adjoints,
but with two additions required by a finite sequence change: all affected
primitive factors are replaced jointly, and mutation-sensitive boundary
contexts are carried explicitly.

\subsection{Context-indexed finite recombination in Rchange}
\label{sec:rchange-relation}

Rchange realizes this principle for nearest-neighbor RNA energy models by
reusing context-indexed inside and outside quantities
\citep{Kiryu2012}. For an RNA dynamic-programming item \(u\) of state
type \(\tau\), write
\begin{equation}
  Z_u^\tau\!\left(
    \kappa^{\mathrm{in}};x_{-p}
  \right),
  \qquad
  W_u^\tau\!\left(
    \kappa^{\mathrm{out}};x_{-p}
  \right)
  \label{eq:rna-context-indexed-items}
\end{equation}
for reusable inside and outside quantities indexed by the boundary
contexts required by the energy model. Here \(x_{-p}\) denotes the
unchanged sequence background, with the nucleotide at position \(p\)
supplied through the exposed local factors and context arguments.

Let \(\gamma\) index the mutually exclusive recombination cases over the
relevant state types, interval items, loop classes, split points, and
boundary configurations. For each case, let
\(\Phi_\gamma(c;\kappa_\gamma(c))\) be the joint product of all primitive
Boltzmann factors that depend on the candidate nucleotide \(c\). The
Rchange-type reconstruction can then be written schematically as
\begin{align}
  R^{\mathrm{Rch}}_{p,x}(c)
  &=
  \sum_{\tau,u,\gamma}
  W_u^\tau\!\left(
    \kappa_\gamma^{\mathrm{out}}(c);x_{-p}
  \right)
  \Phi_\gamma\!\left(
    c;\kappa_\gamma(c)
  \right)
  \nonumber\\
  &\qquad\qquad\times
  \prod_{v\in\operatorname{ch}(\gamma)}
  Z_{u_{\gamma,v}}^{\tau_{\gamma,v}}\!\left(
    \kappa_{\gamma,v}^{\mathrm{in}}(c);x_{-p}
  \right).
  \label{eq:rna-context-recombination}
\end{align}
The notation suppresses implementation-specific state bookkeeping; its
essential content is Eq.~\eqref{eq:rna-recombination-principle}. When the
case enumeration is complete and nonredundant,
\begin{equation}
  Z_{\mathrm{RNA}}\!\left(
    x^{p\to c}
  \right)
  =
  R^{\mathrm{Rch}}_{p,x}(c),
  \qquad
  Z_{\mathrm{RNA}}(x)
  =
  R^{\mathrm{Rch}}_{p,x}(x_p).
  \label{eq:rna-context-recombination-endpoints}
\end{equation}
Thus Rchange evaluates a finite mutant endpoint, not a first-order
linearization around the reference sequence.

The connection to Section~2 is particularly simple in the
context-independent limit. If the boundary contexts do not depend on
\(c\), and each case exposes one coordinate from one affine
mutation-dependent factor group, the inside and outside pieces become
fixed derivative coefficients and
Eq.~\eqref{eq:rna-context-recombination} reduces to the exact one-group
replacement formula. Rchange-type recombination can therefore be viewed as
a context-dependent extension of the same finite factorization rather than
as a different differential principle.

Equation~\eqref{eq:rna-context-recombination} is intentionally schematic.
Exactness for a full Turner-style implementation depends on a complete and
nonredundant enumeration of its state, loop, and boundary-context cases.
Supplementary Section~S4 gives the boundary-context construction, an
explicit four-role RNA-like SCFG specialization, and the double-mutant
background expansion. The full Rfold/Turner case table remains
implementation-dependent and is not reproduced here.

\subsection{Multiple mutations and finite interaction}
\label{sec:rna-multiple-mutations}

For two substitutions \(x_p\to c\) and \(x_q\to d\), let
\(x'=x^{q\to d}\) be the background containing the first edit. Exact
evaluation of the second edit uses inside and outside quantities appropriate
to this mutant background:
\begin{equation}
  Z_{\mathrm{RNA}}\!\left(
    x^{p\to c,q\to d}
  \right)
  =
  R^{\mathrm{Rch}}_{p,x'}(c).
  \label{eq:rna-double-mutant-recombination}
\end{equation}
The mixed finite interaction of Section~2 can therefore be written as the
change in the \(p\)-mutation effect between the two backgrounds:
\begin{align}
  \Delta_p\Delta_q Z_{\mathrm{RNA}}
  &=
  \left[
    R^{\mathrm{Rch}}_{p,x'}(c)
    -
    R^{\mathrm{Rch}}_{p,x'}(x_p)
  \right]
  \nonumber\\
  &\quad-
  \left[
    R^{\mathrm{Rch}}_{p,x}(c)
    -
    R^{\mathrm{Rch}}_{p,x}(x_p)
  \right].
  \label{eq:rna-interaction-as-background-change}
\end{align}
The interaction can arise because both positions affect the same local
thermodynamic factor or because one mutation changes the boundary context
in which the other is evaluated. Applying the edits in the opposite order
gives a different computational decomposition of the same double-mutant
endpoint. The algebraic decomposition into the two one-site effects and
their mixed finite difference remains the one given in Section~2; what
changes in RNA is how each endpoint is reconstructed.

\FloatBarrier
\section{Discussion}
\label{sec:discussion}

The framework separates several operations that are often discussed
together: posterior inference, expected-count learning, infinitesimal
sensitivity, and exact finite sequence replacement. The main practical
consequence is that forward/inside and backward/outside quantities can often
be reused across many candidate edits, but the exact form of that reuse is
determined by the sequence-to-factor map.

\subsection{Computational implications}
\label{sec:discussion-computational-implications}

The finite-replacement identities are most useful when many substitutions
are scored against one reference sequence. After one forward/inside and one
backward/outside computation, HMM, SCFG, and alignment candidates reduce to
contractions of stored coefficients with the factors changed by the edit.
For nearest-neighbor RNA, the analogous reuse is context-dependent rather
than a single affine contraction. Representative costs are summarized in
Table~\ref{tab:discussion-complexity}.

\begin{table}[t]
  \centering
  \small
  \renewcommand{\arraystretch}{1.2}
  \caption{
    Representative computational costs after one forward/inside and one
    backward/outside computation. Per-candidate costs exclude the initial
    dynamic-programming pass. The Rchange entry is the published
    span-limited complexity for a fixed four-letter RNA alphabet and
    maximum base-pair span \(W\); \(O(W^2)\) is an amortized cost within the
    exhaustive single-mutant scan.
  }
  \label{tab:discussion-complexity}
  \begin{tabular}{
    >{\raggedright\arraybackslash}p{0.18\textwidth}
    >{\raggedright\arraybackslash}p{0.23\textwidth}
    >{\raggedright\arraybackslash}p{0.20\textwidth}
    >{\raggedright\arraybackslash}p{0.23\textwidth}
  }
    \hline
    Model
    &
    Base forward/inside--outside cost
    &
    One substitution candidate
    &
    All one-site candidates
    \\
    \hline
    HMM
    &
    \(O(LQ^2)\)
    &
    \(O(Q)\)
    &
    \(O(L\lvert\Sigma\rvert Q)\)
    \\
    Affine-gap alignment, substitutions in \(x\)
    &
    \(O(LK)\)
    &
    \(O(K)\)
    &
    \(O(LK\lvert\Sigma\rvert)\)
    \\
    SCFG with binary and lexical rules
    &
    \(O(L^3\lvert R_2\rvert)\)
    &
    \(O(N)\)
    &
    \(O(L\lvert\Sigma\rvert N)\)
    \\
    Rchange, nearest-neighbor RNA with maximum span \(W\)
    &
    \(O(LW^2)\)
    &
    \(O(W^2)\) amortized
    &
    \(O(LW^2)\)
    \\
    \hline
  \end{tabular}
\end{table}

The gain is not a reduction of the base dynamic-programming complexity.
Rather, an all-substitution scan replaces repeated global reruns by one
global forward--reverse computation followed by local contractions or
context-dependent recombination. This is the relevant regime for sequence
design, where every alternative symbol at every position may be evaluated
against the same background.

Multisite scans remain more expensive. A multi-affine expansion contains
mixed terms for nonempty subsets of the edited positions, while in RNA a
first mutation can also change the context in which a second is evaluated.
For span-limited Rchange, exhaustive single-mutant analysis costs
\(O(LW^2)\), unrestricted double-mutant analysis costs \(O(L^2W^2)\), and a
maximum mutation separation \(D\) gives \(O(LW^2D)\)
\citep{Kiryu2012}. Reverse-mode automatic differentiation can generate
adjoints, but it does not by itself identify the biologically complete
mutation-dependent factor groups or construct the context enumeration needed
for these exact finite edits.

\subsection{Numerical verification}
\label{sec:discussion-numerical-verification}

We verified the finite identities by comparing formula-based mutant values
with direct reconstruction and complete rerunning of each dynamic program.
The tests cover HMM, affine-gap alignment, binary--lexical SCFG, and the
five-rule RNA-like SCFG used in Supplementary Section~S4. Complete
parameters, random seeds, source code, and additional residuals are given in
Supplementary Section~S5.

For a formula value \(Z_{\mathrm{formula}}\) and a directly recomputed value
\(Z_{\mathrm{direct}}\), define
\begin{equation}
  r_{\mathrm{exact}}
  =
  \frac{
    \left|
      Z_{\mathrm{formula}}-Z_{\mathrm{direct}}
    \right|
  }{
    Z(x)
  }.
  \label{eq:numerical-exact-residual}
\end{equation}
Table~\ref{tab:numerical-verification} shows the maximum residuals in one
fixed instance of each model.

\begin{table}[t]
  \centering
  \small
  \renewcommand{\arraystretch}{1.2}
  \caption{
    Numerical verification of the exact finite-replacement identities.
    Across 50 additional random parameter sets per model, every normalized
    residual remained below \(2\times10^{-14}\).
  }
  \label{tab:numerical-verification}
  \begin{tabular}{
    >{\raggedright\arraybackslash}p{0.31\textwidth}
    >{\raggedright\arraybackslash}p{0.36\textwidth}
    >{\raggedright\arraybackslash}p{0.19\textwidth}
  }
    \hline
    Model
    &
    Verified identity
    &
    Maximum residual
    \\
    \hline
    HMM
    &
    One-site replacement
    &
    \(3\times10^{-16}\)
    \\
    HMM
    &
    Two-site mixed expansion
    &
    \(2\times10^{-15}\)
    \\
    Affine-gap alignment
    &
    One-site replacement
    &
    \(1\times10^{-16}\)
    \\
    SCFG with binary and lexical rules
    &
    One-site replacement
    &
    \(2\times10^{-16}\)
    \\
    RNA-like SCFG
    &
    One-site recombination
    &
    \(8\times10^{-16}\)
    \\
    RNA-like SCFG
    &
    Two-site background recombination
    &
    \(2\times10^{-15}\)
    \\
    \hline
  \end{tabular}
\end{table}

All exact identities agree with direct recomputation at double-precision
round-off. The numerical tests are implementation checks, not proofs; the
exactness follows from the algebraic results of Sections~2.6 and~4.

The same calculations also illustrate what the exact identities do
\emph{not} imply. Omitting the mixed term from a two-site change produced
\begin{equation}
  r_{\mathrm{add}}
  =
  \frac{
    \left|
      \Delta_{p,q}Z
      -
      \left(
        \Delta_pZ+\Delta_qZ
      \right)
    \right|
  }{
    Z(x)
  }
  =
  \frac{
    \left|
      \Delta_p\Delta_qZ
    \right|
  }{
    Z(x)
  },
  \label{eq:numerical-additive-residual}
\end{equation}
with a maximum of \(0.96\) in the fixed HMM instance. Likewise, replacing
the exact finite change of \(\log Z\) by its first-order form,
\begin{equation}
  r_{\log}
  =
  \left|
    \Delta_p\log Z
    -
    \frac{\Delta_pZ}{Z(x)}
  \right|,
  \label{eq:numerical-log-residual}
\end{equation}
gave a maximum of \(0.58\) nats. Exact reconstruction of \(Z\) therefore
does not make an additive multisite approximation or a first-order
transformation of \(Z\) exact.

The RNA rows concern the simplified RNA-like SCFG, not an independent
validation of the complete Turner/Rfold case enumeration used by Rchange.
That distinction is detailed in Supplementary Sections~S4 and~S5.

\subsection{Scope and limitations}
\label{sec:discussion-scope-limitations}

The framework assumes a finite acyclic sum--product computation with an
explicit local factorization. Posterior interpretations require
nonnegative latent-object weights and \(Z>0\), while logarithmic derivatives
also require care at zero-valued factors.

The exact finite-replacement results concern \(Z\), or another linear
functional of the root quantities. Nonlinear derived quantities such as
\(\log Z\), posterior ratios, base-pairing probabilities, and normalized
expectations generally require reconstruction of their numerator and
denominator endpoints rather than one original-background derivative
contraction.

Multi-affinity is a property of the chosen sequence-to-factor
representation, not of a model name. It can fail through jointly used
mutation-dependent factors, nonlinear parameter tying, or nonlocal
couplings.
Insertions and deletions pose a further problem because
they can change the dynamic-programming domain and latent-object universe
rather than only local factor values, and may therefore require an
enlarged state space or a different finite-recombination construction.
For nearest-neighbor RNA, exactness of the
schematic recombination in Section~4 additionally depends on a complete and
nonredundant enumeration of the implementation-specific state and boundary
contexts.

Finally, exactness does not remove combinatorial growth. Mixed terms increase
with the number of edited positions, and context-dependent models may
require intermediate mutant backgrounds. The framework identifies which
quantities can be reused and which interaction terms are required; it does
not make arbitrary high-order perturbation scans inexpensive.

\subsection{Broader perspective}
\label{sec:discussion-broader-perspective}

The adjoint interpretation of backward and outside quantities is
well established. The additional point developed here is that differential
coefficients determine an exact finite sequence endpoint only when the
sequence-to-factor structure permits it. This distinction is particularly
relevant to sequence design.

A gradient can rank edits or guide optimization on a continuous relaxation,
but it is generally only local information. For a continuously
differentiable extension \(F\),
\begin{equation}
  F(x')-F(x)
  =
  \int_0^1
  \nabla F\!\left(
    x+t(x'-x)
  \right)^{\mathsf{T}}
  (x'-x)
  \,\mathrm{d}t.
  \label{eq:discussion-path-integral}
\end{equation}
The gradient at the reference sequence is the tangent at \(t=0\). In the
one-group affine cases with \(F=Z\), the relevant directional derivative is
constant along the replacement segment, so the integral collapses to the
exact finite-replacement identity. For multisite or context-dependent edits,
the missing mixed terms or mutant contexts describe precisely why this
collapse fails.

The same local-versus-endpoint distinction has a max/min analogue.
Reverse propagation through a unique Viterbi, maximum-score, or
minimum-free-energy optimum follows the traceback and returns the feature
vector of that optimum; ties require generalized derivatives. This
connection is summarized in Supplementary Section~S1.9 and is not needed
for the sum--product results developed here.

Taken together, the framework separates two questions that should not be
conflated: what can be obtained by differentiating a fixed dynamic program,
and what is required to evaluate a finite edit of its observed sequence.
Adjoints answer the first question universally within the sum--product
setting. The second is controlled by the sequence-to-factor map:
group-wise affinity gives exact finite contraction, whereas overlapping
context-dependent factors require broader recombination.

\section*{Acknowledgements}
The author thanks Goro Terai and Takumi Otagaki for constructive discussions and
the participants of the RNA Informatics Dojo 2025 in Nagasaki, Japan, for
fruitful discussions.

\section*{Funding}
This work was supported by Japan Society for the Promotion of Science (JSPS)
KAKENHI grant numbers JP24H00737 and JP22H04925 (PAGS), and Japan Science and
Technology Agency (JST) CREST grant number JPMJCR23N1.

\section*{Declaration of conflicting interests}
The author declares no potential conflicts of interest with respect to the
research, authorship, and/or publication of this article.

\section*{Data and Software Availability}
No new biological datasets were generated or analyzed in this theoretical study.
The only software is a short, self-contained verification script
(\texttt{verify\_numerics.py}), provided with the supplementary material, which
reproduces the machine-precision residuals reported in the numerical
verification section.

\section*{Generative AI disclosure}
OpenAI ChatGPT and Anthropic Claude/Claude Code were used to assist with
scientific editing, structural reorganization, drafting and revision of English
and LaTeX exposition, consistency checking, mathematical review, file-level
integration, and compilation. They did not determine the research question or
the final scientific conclusions. The author independently examined all
derivations, numerical results, citations, and wording and takes full
responsibility for the work.

\bibliographystyle{plainnat}
\bibliography{references}
\end{document}


\maketitle
\section*{Contents and relation to the main manuscript}
This Supplementary Material contains the detailed derivations and numerical
checks supporting the compact formulations in the main manuscript.
Section S1 develops the general HMM/SCFG adjoint and finite-replacement theory.
Section S2 gives the full affine-gap alignment derivations.
Section S3 treats RNA ensemble differentiation and coupled-grammar learning at
a fixed sequence. Section S4 develops finite RNA sequence changes and
context-dependent recombination. Section S5 documents numerical verification
and reproducibility, including the Supplementary Code
\texttt{anc/verify\_numerics.py}.

\section{Detailed Adjoint and Finite-Replacement Theory for HMMs and SCFGs}
\label{sec:supp-hmm-scfg-theory}

This section provides the detailed derivations supporting Section~2 of the
main text. We first formalize the common sum--product computation, then derive
the HMM backward and SCFG outside recursions by reverse differentiation. We
next place the conditional latent-variable distribution in exponential-family
form and derive item-use, local-event, and shared-feature expectations from
log-partition derivatives and adjoint contractions. We then establish the
exact one-position substitution formulas for HMMs and SCFGs, explain why
finite replacements are not generally first derivatives, and give a complete
proof of the multi-affine finite-replacement expansion. The final subsection
records the corresponding generalized-derivative interpretation of max/min
dynamic programs and traceback.

\subsection{Finite acyclic sum--product computations}
\label{sec:supp-formal-sum-product}

Let \(\mathcal{G}\) be a finite acyclic computation graph whose nodes are
forward or inside items. For each item \(u\), let \(\mathcal{R}(u)\) be the
finite set of local decompositions that create \(u\). A decomposition
\(r\in\mathcal{R}(u)\) has a nonnegative local factor \(\phi_r\) and an
ordered tuple of child items \(\operatorname{ch}(r)\). Repeated child items
are retained with their multiplicity. The forward or inside recursion is
\begin{equation}
  A_u
  =
  \sum_{r\in\mathcal{R}(u)}
  \phi_r
  \prod_{v\in\operatorname{ch}(r)}
  A_v.
  \label{eq:supp-general-inside}
\end{equation}
Let \(u_{\mathrm{root}}\) be the root item and define
\begin{equation}
  Z
  =
  A_{u_{\mathrm{root}}}.
  \label{eq:supp-root-partition}
\end{equation}
Acyclicity ensures that the items can be evaluated in a topological order and
that repeated substitution of Eq.~\eqref{eq:supp-general-inside} yields a
finite polynomial in the local factors.

A complete recursive choice of local decompositions defines a latent object
\(\omega\in\Omega\). Depending on the model, \(\omega\) may be a hidden-state
path, an alignment path, a derivation tree, or an RNA secondary structure. If
\(N_r(\omega)\) denotes the number of occurrences of decomposition \(r\) in
\(\omega\), then
\begin{equation}
  w(\omega)
  =
  \prod_r
  \phi_r^{N_r(\omega)},
  \label{eq:supp-latent-weight}
\end{equation}
and expansion of the root recursion gives
\begin{equation}
  Z
  =
  \sum_{\omega\in\Omega}
  w(\omega).
  \label{eq:supp-latent-expansion}
\end{equation}
For an ambiguous SCFG, the elements of \(\Omega\) are derivation trees rather
than terminal strings or unlabelled structures, so distinct derivations are
counted separately.

When sequence-dependent admissibility changes under substitution, we compare
the original and mutant sequences on a common finite combinatorial universe
\(\Omega\). A local choice that is incompatible with a particular sequence is
assigned factor zero. Thus the latent-object universe remains fixed while the
sequence changes the coefficients of the polynomial in
Eq.~\eqref{eq:supp-latent-expansion}. Posterior interpretations require
\(Z>0\), but the polynomial and finite-replacement identities remain valid at
individual zero-valued local factors.

\subsection{HMM forward--backward recursions by reverse differentiation}
\label{sec:supp-hmm-adjoints}

Let \(x=x_1\cdots x_L\), let \(\mathcal{S}\) be the hidden-state set, and let
\(\pi_s\), \(a_{st}\), and \(e_s(c)\) denote initial, transition, and
emission probabilities. The path weight is
\begin{equation}
  w_{\mathrm{HMM}}(z;x)
  =
  \pi_{z_1}e_{z_1}(x_1)
  \prod_{i=2}^{L}
  a_{z_{i-1}z_i}e_{z_i}(x_i).
  \label{eq:supp-hmm-path-weight}
\end{equation}
The forward variables satisfy
\begin{align}
  \alpha_1(s)
  &=
  \pi_s e_s(x_1),
  \label{eq:supp-hmm-forward-initial}
  \\
  \alpha_i(t)
  &=
  e_t(x_i)
  \sum_{s\in\mathcal{S}}
  \alpha_{i-1}(s)a_{st},
  \qquad i=2,\ldots,L,
  \label{eq:supp-hmm-forward-recursion}
\end{align}
and
\begin{equation}
  Z_{\mathrm{HMM}}(x)
  =
  \sum_{s\in\mathcal{S}}
  \alpha_L(s).
  \label{eq:supp-hmm-forward-termination}
\end{equation}

Define the adjoint of a forward variable by
\begin{equation}
  \beta_i(s)
  =
  \frac{\partial Z_{\mathrm{HMM}}}{\partial\alpha_i(s)}.
  \label{eq:supp-hmm-adjoint-definition}
\end{equation}
Equation~\eqref{eq:supp-hmm-forward-termination} gives the terminal condition
\begin{equation}
  \beta_L(s)
  =
  1.
  \label{eq:supp-hmm-backward-boundary}
\end{equation}
For \(i<L\), the forward variable \(\alpha_i(s)\) influences every
\(\alpha_{i+1}(t)\), and
\begin{equation}
  \frac{\partial\alpha_{i+1}(t)}{\partial\alpha_i(s)}
  =
  a_{st}e_t(x_{i+1}).
  \label{eq:supp-hmm-local-jacobian}
\end{equation}
The chain rule therefore gives
\begin{align}
  \beta_i(s)
  &=
  \sum_{t\in\mathcal{S}}
  \frac{\partial Z_{\mathrm{HMM}}}
       {\partial\alpha_{i+1}(t)}
  \frac{\partial\alpha_{i+1}(t)}
       {\partial\alpha_i(s)}
  \nonumber\\
  &=
  \sum_{t\in\mathcal{S}}
  a_{st}e_t(x_{i+1})\beta_{i+1}(t),
  \qquad i=L-1,\ldots,1.
  \label{eq:supp-hmm-backward-recursion}
\end{align}
Thus the standard backward recursion is reverse-mode differentiation of the
forward computation.

For position \(i\), define
\begin{equation}
  F_i(s)
  =
  \begin{cases}
    \pi_s,
    & i=1,\\[4pt]
    \displaystyle
    \sum_{r\in\mathcal{S}}
    \alpha_{i-1}(r)a_{rs},
    & i\geq 2.
  \end{cases}
  \label{eq:supp-hmm-prefix-coefficient}
\end{equation}
Then
\begin{equation}
  \alpha_i(s)
  =
  e_s(x_i)F_i(s).
  \label{eq:supp-hmm-forward-factorization}
\end{equation}
If the occurrence-specific emission coordinate at position \(i\) is denoted
by \(\rho^{\mathrm{HMM}}_{i,s}=e_s(x_i)\), the chain rule gives
\begin{equation}
  \frac{\partial Z_{\mathrm{HMM}}}
       {\partial\rho^{\mathrm{HMM}}_{i,s}}
  =
  F_i(s)\beta_i(s).
  \label{eq:supp-hmm-position-factor-derivative}
\end{equation}
Neither \(F_i(s)\) nor \(\beta_i(s)\) depends on any coordinate in the
position-\(i\) emission group: the former uses only the prefix and the latter
only the suffix.

The derivatives with respect to shared HMM parameters are obtained by
accumulating the corresponding local contributions. For a transition
parameter,
\begin{equation}
  \frac{\partial Z_{\mathrm{HMM}}}{\partial a_{st}}
  =
  \sum_{i=2}^{L}
  \alpha_{i-1}(s)e_t(x_i)\beta_i(t).
  \label{eq:supp-hmm-transition-derivative}
\end{equation}
For an initial probability,
\begin{equation}
  \frac{\partial Z_{\mathrm{HMM}}}{\partial\pi_s}
  =
  e_s(x_1)\beta_1(s).
  \label{eq:supp-hmm-initial-derivative}
\end{equation}
For a shared emission parameter \(e_s(c)\),
\begin{equation}
  \frac{\partial Z_{\mathrm{HMM}}}{\partial e_s(c)}
  =
  \sum_{i:\,x_i=c}
  F_i(s)\beta_i(s).
  \label{eq:supp-hmm-emission-derivative}
\end{equation}
When the corresponding parameter is positive, multiplication by the parameter
and division by \(Z_{\mathrm{HMM}}\) gives the posterior expected count of
that transition, initial-state event, or emission event.

\subsection{SCFG inside--outside recursions by reverse differentiation}
\label{sec:supp-scfg-adjoints}

Consider an ordinary yield-generating SCFG in Chomsky normal form for
nonempty strings, with no \(\epsilon\)-productions. Its productions are
binary rules \(A\to BC\), with weights \(t_{A\to BC}\), or lexical rules
\(A\to c\), with factors \(e_A(c)\). Let \(\mathcal{N}\) be the
nonterminal set, \(\mathcal{N}_{\mathrm{lex}}\subseteq\mathcal{N}\) the set
of nonterminals that can emit terminals, and \(S\) the start symbol. The
inside variables satisfy
\begin{align}
  I_A(i,i)
  &=
  e_A(x_i),
  \qquad A\in\mathcal{N}_{\mathrm{lex}},
  \label{eq:supp-scfg-inside-terminal}
  \\
  I_A(i,j)
  &=
  \sum_{A\to BC}
  t_{A\to BC}
  \sum_{k=i}^{j-1}
  I_B(i,k)I_C(k+1,j),
  \qquad i<j,
  \label{eq:supp-scfg-inside-recursion}
\end{align}
and
\begin{equation}
  Z_{\mathrm{SCFG}}(x)
  =
  I_S(1,L).
  \label{eq:supp-scfg-root}
\end{equation}
The formulas extend directly to several lexical rule types or additional
nonterminal states by enlarging the set of local decompositions.

Define the outside adjoint by
\begin{equation}
  O_A(i,j)
  =
  \frac{\partial Z_{\mathrm{SCFG}}}{\partial I_A(i,j)}.
  \label{eq:supp-scfg-outside-definition}
\end{equation}
Initialize all outside values to zero and seed the root by
\begin{equation}
  O_S(1,L)
  =
  1.
  \label{eq:supp-scfg-outside-boundary}
\end{equation}
Process parent spans in reverse topological order. For every rule
\(A\to BC\), parent span \((i,j)\), and split point
\(k\in\{i,\ldots,j-1\}\), reverse differentiation of
Eq.~\eqref{eq:supp-scfg-inside-recursion} adds
\begin{align}
  O_B(i,k)
  &\mathrel{+}=
  O_A(i,j)
  t_{A\to BC}
  I_C(k+1,j),
  \label{eq:supp-scfg-outside-left-update}
  \\
  O_C(k+1,j)
  &\mathrel{+}=
  O_A(i,j)
  t_{A\to BC}
  I_B(i,k).
  \label{eq:supp-scfg-outside-right-update}
\end{align}
Accumulation over all parent spans, rules, and split points gives the complete
outside recursion. Ambiguity causes no difficulty: the inside polynomial sums
all derivation trees, and the reverse pass differentiates that same polynomial.

The derivative with respect to a shared binary-rule weight is
\begin{equation}
  \frac{\partial Z_{\mathrm{SCFG}}}
       {\partial t_{A\to BC}}
  =
  \sum_{1\leq i<j\leq L}
  O_A(i,j)
  \sum_{k=i}^{j-1}
  I_B(i,k)I_C(k+1,j).
  \label{eq:supp-scfg-rule-derivative}
\end{equation}
For a shared lexical emission factor,
\begin{equation}
  \frac{\partial Z_{\mathrm{SCFG}}}{\partial e_A(c)}
  =
  \sum_{i:\,x_i=c}
  O_A(i,i),
  \qquad A\in\mathcal{N}_{\mathrm{lex}}.
  \label{eq:supp-scfg-emission-derivative}
\end{equation}
When the corresponding parameter is positive, its logarithmic derivative is
the expected count of that binary rule or lexical emission.

For an occurrence-specific terminal coordinate
\(\rho^{\mathrm{SCFG}}_{i,A}=e_A(x_i)\),
Eq.~\eqref{eq:supp-scfg-inside-terminal} and the chain rule give
\begin{equation}
  \frac{\partial Z_{\mathrm{SCFG}}}
       {\partial\rho^{\mathrm{SCFG}}_{i,A}}
  =
  O_A(i,i).
  \label{eq:supp-scfg-position-factor-derivative}
\end{equation}
The coefficient \(O_A(i,i)\) is independent of every terminal factor at
position \(i\), because the outside context excludes the selected lexical
item and a complete derivation emits position \(i\) exactly once.

\subsection{General adjoints, exponential-family identities, and expected counts}
\label{sec:supp-general-adjoint-identities}

For the general recursion in Eq.~\eqref{eq:supp-general-inside}, define
\begin{equation}
  B_u
  =
  \frac{\partial Z}{\partial A_u}.
  \label{eq:supp-general-adjoint-definition}
\end{equation}
The root is seeded by
\begin{equation}
  B_{u_{\mathrm{root}}}
  =
  1.
  \label{eq:supp-general-root-seed}
\end{equation}
Every non-root item can influence \(Z\) through every parent item whose
recursion contains it. The multivariable chain rule gives
\begin{equation}
  B_u
  =
  \mathbf{1}\{u=u_{\mathrm{root}}\}
  +
  \sum_{p:\,A_p\text{ depends on }A_u}
  B_p
  \frac{\partial A_p}{\partial A_u}.
  \label{eq:supp-general-reverse-recursion}
\end{equation}
If an item occurs more than once in one parent term, its multiplicity is
included in the local derivative.

For a particular decomposition \(r\in\mathcal{R}(u)\), define
\begin{equation}
  T_r
  =
  \phi_r
  \prod_{v\in\operatorname{ch}(r)}
  A_v.
  \label{eq:supp-general-local-term}
\end{equation}
The reverse contribution from this term to one selected occurrence of child
\(v\) is
\begin{equation}
  B_u
  \phi_r
  \prod_{v'\in\operatorname{ch}(r)\setminus\{v\}_{\mathrm{occ}}}
  A_{v'},
  \label{eq:supp-general-local-reverse-contribution}
\end{equation}
where \(\setminus\{v\}_{\mathrm{occ}}\) means that only the selected
occurrence is omitted. Summing these contributions over all occurrences and
parents gives Eq.~\eqref{eq:supp-general-reverse-recursion}.

For a fixed observed sequence \(x\), the latent distributions of HMMs and
SCFGs have an exponential family form. More generally, collect the positive
local factor types into \(\{\psi_k\}\), and let \(N_k(\omega)\) denote
the multiplicity of factor \(\psi_k\) in latent object \(\omega\). Then
\begin{equation}
  w(\omega)
  =
  \prod_k
  \psi_k^{N_k(\omega)}.
  \label{eq:supp-shared-factorization}
\end{equation}
With the log-factor coordinates
\begin{equation}
  \theta_k
  =
  \log\psi_k,
  \label{eq:supp-log-factor-coordinate}
\end{equation}
the normalized latent distribution is
\begin{equation}
  P(\omega\mid x;\boldsymbol{\theta})
  =
  \frac{w(\omega)}{Z(\boldsymbol{\theta})}
  =
  \exp\!\left\{
    \sum_k
    \theta_kN_k(\omega)
    -
    \log Z(\boldsymbol{\theta})
  \right\}.
  \label{eq:supp-exponential-family}
\end{equation}
Thus the counts \(N_k\) are sufficient statistics, and the standard
log-partition identity gives
\begin{equation}
  \frac{\partial\log Z}{\partial\theta_k}
  =
  \frac{\partial\log Z}{\partial\log\psi_k}
  =
  \frac{\psi_k}{Z}
  \frac{\partial Z}{\partial\psi_k}
  =
  \mathbb{E}\!\left[
    N_k
    \mid
    x
  \right].
  \label{eq:supp-log-derivative-expected-count}
\end{equation}
When HMM probabilities or SCFG rule probabilities are subject to
normalization constraints, Eq.~\eqref{eq:supp-exponential-family} is most
conveniently read in the independent positive factor coordinates. The
constraints restrict the admissible log-factor parameters; derivatives in a
normalized or otherwise reparameterized model follow by the chain rule.

Item use is obtained by introducing an auxiliary exponential-family marker.
Let \(N_u(\omega)\) be the number of occurrences of item \(u\) in the fully
unfolded latent object. Multiply every downstream use of \(A_u\) by a common
marker \(\lambda_u>0\), and set
\begin{equation}
  \eta_u
  =
  \log\lambda_u.
  \label{eq:supp-item-natural-parameter}
\end{equation}
Equivalently, the marked partition function is
\begin{equation}
  Z_u(\lambda_u)
  =
  \sum_{\omega\in\Omega}
  w(\omega)
  \lambda_u^{N_u(\omega)}.
  \label{eq:supp-item-marked-partition}
\end{equation}
For the root item, the marker is applied to the root output itself, so that
\(N_{u_{\mathrm{root}}}(\omega)=1\). From the exponential-family identity,
\begin{equation}
  \left.
  \frac{\partial\log Z_u}{\partial\eta_u}
  \right|_{\eta_u=0}
  =
  \left.
  \frac{\partial\log Z_u}{\partial\log\lambda_u}
  \right|_{\lambda_u=1}
  =
  \mathbb{E}\!\left[
    N_u
    \mid
    x
  \right].
  \label{eq:supp-item-marker-expected-count}
\end{equation}
On the dynamic-programming graph, scaling every downstream use of \(A_u\) by
\(\lambda_u\) gives
\begin{equation}
  \left.
  \frac{\partial Z_u}{\partial\lambda_u}
  \right|_{\lambda_u=1}
  =
  A_u
  \frac{\partial Z}{\partial A_u}
  =
  A_uB_u.
  \label{eq:supp-item-marker-chain-rule}
\end{equation}
Combining the two representations yields the unified identity
\begin{equation}
  \mathbb{E}\!\left[
    N_u
    \mid
    x
  \right]
  =
  \left.
  \frac{\partial\log Z_u}{\partial\eta_u}
  \right|_{\eta_u=0}
  =
  \left.
  \frac{\partial\log Z_u}{\partial\log\lambda_u}
  \right|_{\lambda_u=1}
  =
  \frac{A_u}{Z}
  \frac{\partial Z}{\partial A_u}
  =
  \frac{A_uB_u}{Z}.
  \label{eq:supp-item-use-identity}
\end{equation}
The marker supplies the exponential-family parameter whose sufficient
statistic is item use; \(A_u\) itself is an intermediate dynamic-programming
variable rather than a primitive model parameter. When every latent object
uses item \(u\) at most once, Eq.~\eqref{eq:supp-item-use-identity} is an
occurrence probability; otherwise it is an expected occurrence count.

The same construction applies to one local decomposition. Replace
\(T_r\) in the recursion for \(A_u\) by \(\lambda_rT_r\). Then
\begin{equation}
  \mathbb{E}\!\left[
    N_r
    \mid
    x
  \right]
  =
  \left.
  \frac{\partial\log Z}{\partial\log\lambda_r}
  \right|_{\lambda_r=1}
  =
  \frac{B_uT_r}{Z}.
  \label{eq:supp-local-event-identity}
\end{equation}
Thus the item identity sums over all decompositions that create \(u\), while
the local-event identity selects one decomposition. The HMM state posterior
and SCFG constituent posterior are item quantities; the HMM
transition--emission posterior and SCFG production--split posterior are local
event quantities.


For completeness, direct differentiation of
Eq.~\eqref{eq:supp-shared-factorization} gives
\begin{equation}
  \frac{\partial Z}{\partial\psi_k}
  =
  \sum_{\omega\in\Omega}
  N_k(\omega)
  \psi_k^{N_k(\omega)-1}
  \prod_{\ell\neq k}
  \psi_\ell^{N_\ell(\omega)}.
  \label{eq:supp-shared-factor-derivative}
\end{equation}
Reverse differentiation computes this derivative by accumulating the local
adjoint contribution from every occurrence that uses \(\psi_k\).

If \(\psi_k=0\), the logarithmic coordinates in
Eqs.~\eqref{eq:supp-log-factor-coordinate}--\eqref{eq:supp-log-derivative-expected-count}
are undefined. The weighted polynomial identity
\begin{equation}
  \frac{\psi_k}{Z}
  \frac{\partial Z}{\partial\psi_k}
  =
  \mathbb{E}\!\left[
    N_k
    \mid
    x
  \right]
  \label{eq:supp-zero-factor-identity}
\end{equation}
nevertheless remains valid by direct evaluation when \(Z>0\). Every
positive-weight latent object then has \(N_k(\omega)=0\), so both sides are
zero. Expressions that divide by \(\psi_k\) must not be used at a zero
factor.

\subsection{Exact one-position substitutions in HMMs and SCFGs}
\label{sec:supp-hmm-scfg-one-site}

For the HMM, define the position-specific emission group
\begin{equation}
  \boldsymbol{\rho}^{\mathrm{HMM}}_i
  =
  \left(
    \rho^{\mathrm{HMM}}_{i,s}
  \right)_{s\in\mathcal{S}},
  \qquad
  \rho^{\mathrm{HMM}}_{i,s}
  =
  e_s(x_i).
  \label{eq:supp-hmm-position-group}
\end{equation}
Every hidden-state path emits position \(i\) in exactly one state. Grouping
all path weights by that state gives
\begin{equation}
  Z_{\mathrm{HMM}}(x)
  =
  \sum_{s\in\mathcal{S}}
  F_i(s)\beta_i(s)
  \rho^{\mathrm{HMM}}_{i,s}.
  \label{eq:supp-hmm-position-affine-decomposition}
\end{equation}
The coefficient of each coordinate is independent of the whole group. Hence
for a replacement \(x_i\to c\),
\begin{align}
  Z_{\mathrm{HMM}}\!\left(x^{i\to c}\right)
  -
  Z_{\mathrm{HMM}}(x)
  &=
  \sum_{s\in\mathcal{S}}
  F_i(s)\beta_i(s)
  \left[
    e_s(c)-e_s(x_i)
  \right]
  \nonumber\\
  &=
  \sum_{s\in\mathcal{S}}
  \frac{\partial Z_{\mathrm{HMM}}}
       {\partial\rho^{\mathrm{HMM}}_{i,s}}
  \Delta\rho^{\mathrm{HMM}}_{i,s}.
  \label{eq:supp-hmm-exact-one-site}
\end{align}
No approximation is involved.

For the SCFG, define the position-specific lexical group
\begin{equation}
  \boldsymbol{\rho}^{\mathrm{SCFG}}_i
  =
  \left(
    \rho^{\mathrm{SCFG}}_{i,A}
  \right)_{A\in\mathcal{N}_{\mathrm{lex}}},
  \qquad
  \rho^{\mathrm{SCFG}}_{i,A}
  =
  e_A(x_i).
  \label{eq:supp-scfg-position-group}
\end{equation}
Under the grammar assumptions stated above, every complete derivation emits
position \(i\) through exactly one lexical event. Grouping derivations by the
lexical nonterminal used at that position gives
\begin{equation}
  Z_{\mathrm{SCFG}}(x)
  =
  \sum_{A\in\mathcal{N}_{\mathrm{lex}}}
  O_A(i,i)
  \rho^{\mathrm{SCFG}}_{i,A}.
  \label{eq:supp-scfg-position-affine-decomposition}
\end{equation}
Therefore,
\begin{align}
  Z_{\mathrm{SCFG}}\!\left(x^{i\to c}\right)
  -
  Z_{\mathrm{SCFG}}(x)
  &=
  \sum_{A\in\mathcal{N}_{\mathrm{lex}}}
  O_A(i,i)
  \left[
    e_A(c)-e_A(x_i)
  \right]
  \nonumber\\
  &=
  \sum_{A\in\mathcal{N}_{\mathrm{lex}}}
  \frac{\partial Z_{\mathrm{SCFG}}}
       {\partial\rho^{\mathrm{SCFG}}_{i,A}}
  \Delta\rho^{\mathrm{SCFG}}_{i,A}.
  \label{eq:supp-scfg-exact-one-site}
\end{align}

The HMM and SCFG formulas have the same algebraic explanation. Each complete
latent object uses exactly one coordinate from every position-specific
emission group. The HMM partition function and the SCFG partition function
are therefore homogeneous of degree one in each position group and
multi-affine across positions. Ambiguity of the SCFG does not affect this
conclusion because the exactly-once statement applies separately to every
derivation tree.

The exactness in both models is therefore an algebraic consequence of
linearity in the complete position-specific factor group; it does not rely on
the substitution being small.

\subsection{Finite replacements, factor groups, affinity, and the exactly-once condition}
\label{sec:supp-factor-groups}

A finite sequence replacement is not, in general, equal to a first
derivative evaluated at the original sequence. For a scalar factor
\(\rho\), the derivative gives the tangent contribution
\begin{equation}
  \frac{\partial Z}{\partial\rho}\Delta\rho,
  \label{eq:supp-general-first-order-contraction}
\end{equation}
whereas the finite endpoint difference is
\begin{equation}
  Z(\rho+\Delta\rho)-Z(\rho).
  \label{eq:supp-general-finite-endpoint}
\end{equation}
For example, if \(Z(\rho)=\rho^2\), then
\begin{equation}
  Z(\rho+\Delta\rho)-Z(\rho)
  =
  2\rho\Delta\rho
  +
  (\Delta\rho)^2,
  \label{eq:supp-quadratic-counterexample}
\end{equation}
so the original-point derivative misses the quadratic term. In a latent
sum--product model, such terms arise when several factors controlled by the
same sequence edit can occur jointly in one latent object. Exactness
therefore requires an algebraic restriction on the dependence of \(Z\) on
the complete set of mutation-dependent factors.

Let \(\mathcal{P}\) be a finite set of affected sequence positions. For each
\(p\in\mathcal{P}\), define a finite factor group
\begin{equation}
  \boldsymbol{\rho}_p
  =
  \left(
    \rho_{p,a}
  \right)_{a\in\mathcal{A}_p}.
  \label{eq:supp-general-factor-group}
\end{equation}
With all variables outside group \(p\) fixed, \(Z\) is affine in
\(\boldsymbol{\rho}_p\) if
\begin{equation}
  Z
  =
  C_p
  +
  \sum_{a\in\mathcal{A}_p}
  C_{p,a}\rho_{p,a},
  \label{eq:supp-group-affine}
\end{equation}
where \(C_p\) and \(C_{p,a}\) are independent of
\(\boldsymbol{\rho}_p\). For an affine group,
\begin{equation}
  Z(\boldsymbol{\rho}_p+\Delta\boldsymbol{\rho}_p)
  -
  Z(\boldsymbol{\rho}_p)
  =
  \sum_{a\in\mathcal{A}_p}
  \frac{\partial Z}{\partial\rho_{p,a}}
  \Delta\rho_{p,a},
  \label{eq:supp-affine-one-group-exact}
\end{equation}
so the first derivative contraction is exact for an arbitrary finite
replacement of that group.

The dependence is homogeneous of degree one if \(C_p=0\). Equivalently,
with all other groups fixed,
\begin{equation}
  Z(\ldots,\lambda\boldsymbol{\rho}_p,\ldots)
  =
  \lambda
  Z(\ldots,\boldsymbol{\rho}_p,\ldots).
  \label{eq:supp-group-homogeneity-scaling}
\end{equation}
Homogeneity is stronger than the affinity required for
Eq.~\eqref{eq:supp-affine-one-group-exact}. The function \(Z\) is
multi-affine in the groups
\(\{\boldsymbol{\rho}_p:p\in\mathcal{P}\}\) if it is affine in each group
with the other groups fixed.

Because \(Z\) is a finite polynomial, multi-affinity is equivalent to the
statement that no monomial contains more than one coordinate from the same
group after like terms have been collected. Homogeneity of degree one in
group \(p\) means that every monomial contains exactly one coordinate from
that group. Affine but non-homogeneous dependence permits monomials containing
none of its coordinates.

\begin{proposition}[Structural condition for group-wise affinity]
\label{prop:supp-structural-affinity}
If every latent object uses at most one coordinate from
\(\boldsymbol{\rho}_p\), then \(Z\) is affine in that group. If every latent
object uses exactly one coordinate from the group, then the dependence is
homogeneous of degree one.
\end{proposition}

\begin{proof}
Each latent-object weight is a monomial. Under the at-most-once condition, its
degree in the coordinates of group \(p\) is either zero or one. Summing such
monomials gives Eq.~\eqref{eq:supp-group-affine}. Under the exactly-once
condition, every monomial has degree one in the group, so the constant term
\(C_p\) vanishes.
\end{proof}

The proposition gives a transparent sufficient condition. For arbitrary
algebraic representations, affinity could also arise through cancellation or
redundancy. In the nonnegative latent-object expansions considered here,
there is no cancellation between positive monomials, and the structural
condition directly describes the factor use of the model. Multi-affinity is
nevertheless a property of the chosen sequence-to-factor map, not merely of
the dynamic-programming topology. A biological substitution can coordinate
changes in several primitive factors even when the polynomial is separately
linear in each primitive occurrence.

The HMM and SCFG position groups satisfy the exactly-once condition. An
affine-gap alignment position group generally satisfies only the at-most-once
condition, because a residue can instead be aligned to a gap. A
nearest-neighbor RNA nucleotide can affect several factors occurring jointly
in one structure and therefore does not generally fit either simple
position-group condition.

\subsection{Complete proof of the multi-affine finite-replacement expansion}
\label{sec:supp-multi-affine-proof}

For every \(p\in\mathcal{P}\), let
\begin{equation}
  \boldsymbol{\rho}'_p
  =
  \boldsymbol{\rho}_p
  +
  \Delta\boldsymbol{\rho}_p
  \label{eq:supp-group-replacement}
\end{equation}
and define the directional differential operator
\begin{equation}
  \mathcal{D}_p
  =
  \sum_{a\in\mathcal{A}_p}
  \Delta\rho_{p,a}
  \frac{\partial}{\partial\rho_{p,a}}.
  \label{eq:supp-directional-operator}
\end{equation}
If \(Z\) is affine in group \(p\), translation of that entire group is exact
after one directional derivative:
\begin{equation}
  Z\!\left(
    \boldsymbol{\rho}_p
    +
    \Delta\boldsymbol{\rho}_p
  \right)
  =
  \left(
    I+\mathcal{D}_p
  \right)Z.
  \label{eq:supp-one-group-translation}
\end{equation}
Every second derivative involving two coordinates from the same group is zero,
so there is no remainder term.

Translations in distinct coordinate groups commute. Simultaneous replacement
of all groups in \(\mathcal{P}\) therefore gives
\begin{equation}
  Z(\boldsymbol{\rho}')
  =
  \prod_{p\in\mathcal{P}}
  \left(
    I+\mathcal{D}_p
  \right)
  Z(\boldsymbol{\rho}).
  \label{eq:supp-product-translation}
\end{equation}
Expanding the operator product yields
\begin{equation}
  \prod_{p\in\mathcal{P}}
  \left(
    I+\mathcal{D}_p
  \right)
  =
  \sum_{U\subseteq\mathcal{P}}
  \prod_{p\in U}
  \mathcal{D}_p.
  \label{eq:supp-operator-subset-expansion}
\end{equation}
For \(U\subseteq\mathcal{P}\), define
\begin{equation}
  \mathcal{A}_U
  =
  \prod_{p\in U}
  \mathcal{A}_p,
  \qquad
  \boldsymbol{a}_U
  =
  \left(
    a_p
  \right)_{p\in U}.
  \label{eq:supp-multi-index-set}
\end{equation}
For nonempty \(U\),
\begin{equation}
  \prod_{p\in U}
  \mathcal{D}_p Z
  =
  \sum_{\boldsymbol{a}_U\in\mathcal{A}_U}
  \frac{
    \partial^{\lvert U\rvert}Z
  }{
    \displaystyle
    \prod_{p\in U}
    \partial\rho_{p,a_p}
  }
  \prod_{p\in U}
  \Delta\rho_{p,a_p}.
  \label{eq:supp-mixed-directional-term}
\end{equation}
The empty-subset term is the identity operator and returns the original
partition function. Subtracting it proves
\begin{align}
  Z(\boldsymbol{\rho}')
  -
  Z(\boldsymbol{\rho})
  & =
  \sum_{\emptyset\neq U\subseteq\mathcal{P}}
  \;
  \sum_{\boldsymbol{a}_U\in\mathcal{A}_U}
  \frac{
    \partial^{\lvert U\rvert}Z
  }{
    \displaystyle
    \prod_{p\in U}
    \partial\rho_{p,a_p}
  }
  \prod_{p\in U}
  \Delta\rho_{p,a_p}.
  \label{eq:supp-complete-multi-affine-expansion}
\end{align}
All derivatives are evaluated at the original factors. Since \(Z\) is a
polynomial, mixed derivatives with respect to distinct groups commute. This
is a finite polynomial identity, not a truncated Taylor expansion.

For one group,
\begin{equation}
  \Delta_p Z
  =
  \sum_{a\in\mathcal{A}_p}
  \frac{\partial Z}{\partial\rho_{p,a}}
  \Delta\rho_{p,a}.
  \label{eq:supp-general-one-site}
\end{equation}
For two distinct groups \(p\) and \(q\),
\begin{align}
  \Delta_{p,q}Z
  & =
  \sum_{a\in\mathcal{A}_p}
  \frac{\partial Z}{\partial\rho_{p,a}}
  \Delta\rho_{p,a}
  +
  \sum_{b\in\mathcal{A}_q}
  \frac{\partial Z}{\partial\rho_{q,b}}
  \Delta\rho_{q,b}
  \nonumber\\
  &\quad+
  \sum_{a\in\mathcal{A}_p}
  \sum_{b\in\mathcal{A}_q}
  \frac{
    \partial^2 Z
  }{
    \partial\rho_{p,a}\,
    \partial\rho_{q,b}
  }
  \Delta\rho_{p,a}
  \Delta\rho_{q,b}.
  \label{eq:supp-general-two-site}
\end{align}
For three groups, the expansion contains three one-group terms, three
pairwise mixed terms, and one third-order mixed term. In general there are
\(2^{\lvert\mathcal{P}\rvert}-1\) nonempty subset classes, although
model-specific sparsity can make many coefficients zero or admit a more
efficient recombination.

\subsection{Finite interactions, nonlinear transformations, and limitations}
\label{sec:supp-finite-interactions-limitations}

Let \(\mathsf{T}_p\) denote replacement of group \(p\), and define
\begin{equation}
  \Delta_p
  =
  \mathsf{T}_p-I.
  \label{eq:supp-finite-difference-operator}
\end{equation}
For two distinct groups,
\begin{align}
  \Delta_p\Delta_q Z
  & =
  \left(
    \mathsf{T}_p-I
  \right)
  \left(
    \mathsf{T}_q-I
  \right)Z
  \nonumber\\
  & =
  Z_{pq}-Z_p-Z_q+Z.
  \label{eq:supp-mixed-inclusion-exclusion}
\end{align}
Under multi-affinity,
\begin{equation}
  \Delta_p\Delta_q Z
  =
  \sum_{a\in\mathcal{A}_p}
  \sum_{b\in\mathcal{A}_q}
  \frac{
    \partial^2 Z
  }{
    \partial\rho_{p,a}\,
    \partial\rho_{q,b}
  }
  \Delta\rho_{p,a}
  \Delta\rho_{q,b}.
  \label{eq:supp-mixed-derivative-interaction}
\end{equation}
The total two-group change is
\begin{equation}
  \Delta_{p,q}Z
  =
  \Delta_p Z
  +
  \Delta_q Z
  +
  \Delta_p\Delta_q Z.
  \label{eq:supp-total-two-site-change}
\end{equation}
Equivalently,
\begin{equation}
  \Delta_p\Delta_q Z
  =
  \left.
  \Delta_p Z
  \right|_{q\text{-mutant background}}
  -
  \left.
  \Delta_p Z
  \right|_{\text{original background}}.
  \label{eq:supp-background-dependence}
\end{equation}
Thus the mixed finite difference measures the background dependence of one
finite replacement on the other. Its sign compares signed effects and its
magnitude depends on the scale on which \(Z\) is reported.

Repeated edits at the same sequence position do not define distinct factor
groups. They compose into one final replacement of that position group, so
the subset expansion is indexed by distinct affected positions.

The exact expansion concerns \(Z\), not a nonlinear transformation of \(Z\).
For a one-group replacement,
\begin{equation}
  \Delta_p\log Z
  =
  \log\!\left(
    1+
    \frac{\Delta_p Z}{Z}
  \right).
  \label{eq:supp-log-finite-change}
\end{equation}
The approximation
\begin{equation}
  \Delta_p\log Z
  \approx
  \frac{\Delta_p Z}{Z}
  \label{eq:supp-log-linearization}
\end{equation}
is accurate only when \(\lvert\Delta_p Z/Z\rvert\) is sufficiently small.
Similarly, a normalized posterior
\begin{equation}
  P(E\mid x)
  =
  \frac{Z_E}{Z}
  \label{eq:supp-posterior-ratio}
\end{equation}
need not be affine under replacement even if both numerator and denominator
are separately multi-affine.

Multi-affinity is factorization-dependent. It can fail when one sequence
position changes several primitive factors that occur jointly in one latent
object, when local factors are tied by a nonlinear parameterization, or when
inside and outside cells depend on mutation-sensitive boundary contexts.
Introducing occurrence-specific factors can expose a larger polynomial, but
a biological substitution may still coordinate several factor replacements.
Summing their separate original-background effects then omits joint products.

The multi-affinity condition is sufficient for exactness for every
replacement in the chosen groups. It is not necessary for a particular
increment: special directions can make higher-order terms vanish even when
the full group-wise condition fails. Such accidental simplifications do not
provide a model-wide guarantee.

The framework assumes a finite acyclic computation and a common latent-object
universe. Posterior interpretations additionally require nonnegative weights
and \(Z>0\). Insertions, deletions that change the object universe, cyclic
grammars, pseudoknots, tertiary interactions, and black-box nonlocal
potentials may require an enlarged state space or a different finite
recombination construction.

\subsection{Max/min dynamic programs, generalized derivatives, and traceback}
\label{sec:supp-max-min-traceback}

The main analysis concerns sum--product partition functions, but biological
sequence analysis also uses max-plus and min-plus dynamic programs, including
Viterbi decoding, maximum-score alignment, and minimum-free-energy RNA
folding. Semiring formulations place these recursions in a common algebraic
setting \citep{Goodman1999}.

Let each latent object have an additive score
\begin{equation}
  S(\omega;\boldsymbol{s})
  =
  \sum_r
  s_rN_r(\omega),
  \label{eq:supp-additive-score}
\end{equation}
and define
\begin{equation}
  V(\boldsymbol{s})
  =
  \max_{\omega\in\Omega}
  S(\omega;\boldsymbol{s}).
  \label{eq:supp-max-value}
\end{equation}
Since \(V\) is the maximum of finitely many affine functions, it is convex
and piecewise affine. If the optimum is unique at \(\boldsymbol{s}\), with
optimizer \(\omega^\ast\), then
\begin{equation}
  \frac{\partial V}{\partial s_r}
  =
  N_r(\omega^\ast).
  \label{eq:supp-unique-optimum-gradient}
\end{equation}
The gradient is therefore the feature vector of the selected optimal path,
alignment, derivation, or structure.

At a tie, \(V\) need not be differentiable. Its subdifferential is
\begin{equation}
  \partial V(\boldsymbol{s})
  =
  \operatorname{conv}
  \left\{
    \boldsymbol{N}(\omega)
    :
    \omega\in
    \operatorname*{arg\,max}_{\omega'\in\Omega}
    S(\omega';\boldsymbol{s})
  \right\},
  \label{eq:supp-max-subdifferential}
\end{equation}
where \(\operatorname{conv}\) denotes the convex hull. A deterministic
tie-breaking traceback selects one optimal object and therefore one valid
subgradient represented by that object's feature vector. It is an extreme
subgradient when that feature vector is an extreme point of the convex hull.
Generalized derivatives and smoothed dynamic programs are discussed by
\citet{MenschBlondel2018}.

At the dynamic-program level, products of weights become sums of scores and
sum operations become maxima. Reverse propagation follows the selected
maximizing branches. When the maximizing branch is unique at every visited
item, the reverse pass marks exactly the local choices in the unique
traceback. At ties, the reverse object is set-valued unless a tie-breaking or
smoothing rule is specified.

The sum--product and max-plus views are connected by the temperature-scaled
log-partition function
\begin{equation}
  F_T(\boldsymbol{s})
  =
  T
  \log
  \sum_{\omega\in\Omega}
  \exp\!\left(
    \frac{S(\omega;\boldsymbol{s})}{T}
  \right),
  \qquad T>0.
  \label{eq:supp-temperature-log-partition}
\end{equation}
Its gradient is the Boltzmann expected feature vector:
\begin{equation}
  \frac{\partial F_T}{\partial s_r}
  =
  \mathbb{E}_T[N_r].
  \label{eq:supp-temperature-gradient}
\end{equation}
Moreover,
\begin{equation}
  \lim_{T\to 0^+}
  F_T(\boldsymbol{s})
  =
  V(\boldsymbol{s}).
  \label{eq:supp-zero-temperature-limit}
\end{equation}
If the optimum is unique, the expected feature vector converges to the
feature vector of that optimum. At a tie, limiting gradients lie in the
convex hull of the tied optimum feature vectors.

For a min-plus problem such as minimum-free-energy folding, let
\begin{equation}
  E_{\min}(\boldsymbol{\theta})
  =
  \min_{\sigma\in\Omega}
  \sum_k
  \theta_k f_k(\sigma).
  \label{eq:supp-min-energy-value}
\end{equation}
At a unique minimum \(\sigma^\ast\),
\begin{equation}
  \frac{\partial E_{\min}}{\partial\theta_k}
  =
  f_k(\sigma^\ast).
  \label{eq:supp-min-energy-gradient}
\end{equation}

This optimization-side differential interpretation does not imply that the
gradient at the original sequence gives an exact endpoint effect for an
arbitrary finite edit. Within a region where the same optimum remains active,
the optimal value is affine and the gradient gives the exact score change.
If an edit changes the optimal path or structure, the perturbation crosses a
boundary between affine regions, and the original gradient or traceback alone
does not determine the endpoint change. This is the max/min counterpart of
the distinction between local differential information and exact finite
sequence replacement.

\section{Affine-gap alignment as a specialization of the adjoint and finite-replacement framework}
\label{sec:supp-alignment}

This section gives the complete alignment specialization supporting
Section~3.1 of the main text. We use the same three-state affine-gap
convention as in the main text: \(M\) consumes one symbol from each
sequence, \(X\) consumes one symbol from \(x\) and aligns it to a gap, and
\(Y\) consumes one symbol from \(y\) and aligns it to a gap. Direct
\(X\leftrightarrow Y\) transitions are not included in this convention.
We first state the full boundary conditions and reverse recursions, then
derive local-event posterior probabilities and expected counts. We finally
give the exact one- and multisite sequence-replacement formulas and clarify
their dependence on the chosen alignment parameterization.

\subsection{Forward recursions and boundary conditions}
\label{sec:supp-alignment-forward}

Let \(x=x_1\cdots x_L\) and \(y=y_1\cdots y_K\). Define the positive
Boltzmann-like factors
\begin{equation}
  q(a,b)
  =
  \exp\!\left\{
    \frac{s(a,b)}{\tau}
  \right\},
  \qquad
  p_d
  =
  \exp\!\left\{
    -\frac{d}{\tau}
  \right\},
  \qquad
  p_e
  =
  \exp\!\left\{
    -\frac{e}{\tau}
  \right\},
  \label{eq:supp-align-factors}
\end{equation}
where \(s(a,b)\) is a substitution score, \(d\) and \(e\) are gap-opening
and gap-extension penalties, and \(\tau>0\) is a formal scale parameter.

Let \(F_M(i,j)\), \(F_X(i,j)\), and \(F_Y(i,j)\) denote the total weights
of partial alignment paths reaching lattice point \((i,j)\) in states
\(M\), \(X\), and \(Y\), respectively. For \(i\ge 1\) and \(j\ge 1\),
\begin{align}
  F_M(i,j)
  &=
  q(x_i,y_j)
  \left[
    F_M(i-1,j-1)
    +
    F_X(i-1,j-1)
    +
    F_Y(i-1,j-1)
  \right],
  \label{eq:supp-align-forward-m}
  \\
  F_X(i,j)
  &=
  p_dF_M(i-1,j)
  +
  p_eF_X(i-1,j),
  \label{eq:supp-align-forward-x}
  \\
  F_Y(i,j)
  &=
  p_dF_M(i,j-1)
  +
  p_eF_Y(i,j-1).
  \label{eq:supp-align-forward-y}
\end{align}
A convenient global initialization is
\begin{equation}
  F_M(0,0)=1,
  \qquad
  F_X(0,0)=F_Y(0,0)=0.
  \label{eq:supp-align-origin}
\end{equation}
On the horizontal and vertical boundaries,
\begin{align}
  F_X(i,0)
  &=
  p_dp_e^{\,i-1},
  && i=1,\ldots,L,
  \label{eq:supp-align-boundary-x}
  \\
  F_Y(0,j)
  &=
  p_dp_e^{\,j-1},
  && j=1,\ldots,K,
  \label{eq:supp-align-boundary-y}
\end{align}
with
\begin{equation}
  F_M(i,0)=F_Y(i,0)=0
  \quad (i>0),
  \qquad
  F_M(0,j)=F_X(0,j)=0
  \quad (j>0).
  \label{eq:supp-align-zero-boundaries}
\end{equation}
Equivalently, the boundary values in
Eqs.~\eqref{eq:supp-align-boundary-x}--\eqref{eq:supp-align-boundary-y}
can be generated by the same \(X\)- and \(Y\)-recursions if out-of-domain
terms are interpreted as zero.

The partition function is
\begin{equation}
  Z_{\mathrm{aln}}(x,y)
  =
  F_M(L,K)
  +
  F_X(L,K)
  +
  F_Y(L,K).
  \label{eq:supp-align-partition}
\end{equation}
Thus an alignment path of gap length \(r\ge 1\) contributes the gap factor
\(p_dp_e^{r-1}\). The terminal sum in
Eq.~\eqref{eq:supp-align-partition} permits an alignment to end in any of
the three states.

\subsection{Reverse adjoints and backward boundary conditions}
\label{sec:supp-alignment-adjoints}

Define
\begin{equation}
  B_U(i,j)
  =
  \frac{
    \partial Z_{\mathrm{aln}}
  }{
    \partial F_U(i,j)
  },
  \qquad
  U\in\{M,X,Y\}.
  \label{eq:supp-align-adjoint-definition}
\end{equation}
Because the root quantity is the sum of the three terminal items, the
reverse sweep is seeded by
\begin{equation}
  B_M(L,K)
  =
  B_X(L,K)
  =
  B_Y(L,K)
  =
  1.
  \label{eq:supp-align-terminal-adjoints}
\end{equation}
At an arbitrary lattice point, the complete reverse recursions are most
compactly written with indicator functions:
\begin{align}
  B_M(i,j)
  &=
  \mathbf{1}\{i=L,\ j=K\}
  \nonumber\\
  &\quad+
  \mathbf{1}\{i<L,\ j<K\}
  q(x_{i+1},y_{j+1})
  B_M(i+1,j+1)
  \nonumber\\
  &\quad+
  \mathbf{1}\{i<L\}
  p_dB_X(i+1,j)
  +
  \mathbf{1}\{j<K\}
  p_dB_Y(i,j+1),
  \label{eq:supp-align-backward-m}
  \\
  B_X(i,j)
  &=
  \mathbf{1}\{i=L,\ j=K\}
  \nonumber\\
  &\quad+
  \mathbf{1}\{i<L,\ j<K\}
  q(x_{i+1},y_{j+1})
  B_M(i+1,j+1)
  +
  \mathbf{1}\{i<L\}
  p_eB_X(i+1,j),
  \label{eq:supp-align-backward-x}
  \\
  B_Y(i,j)
  &=
  \mathbf{1}\{i=L,\ j=K\}
  \nonumber\\
  &\quad+
  \mathbf{1}\{i<L,\ j<K\}
  q(x_{i+1},y_{j+1})
  B_M(i+1,j+1)
  +
  \mathbf{1}\{j<K\}
  p_eB_Y(i,j+1).
  \label{eq:supp-align-backward-y}
\end{align}
These equations reduce to the interior recursions shown in the main text
when all indicated successor cells are in range.

The same reverse calculation can be implemented as local accumulation
updates. For each \(M\)-cell with \(i\ge1\) and \(j\ge1\),
\begin{equation}
  B_U(i-1,j-1)
  \mathrel{+}=
  q(x_i,y_j)B_M(i,j),
  \qquad
  U\in\{M,X,Y\}.
  \label{eq:supp-align-reverse-update-m}
\end{equation}
For each \(X\)-cell with \(i\ge1\),
\begin{align}
  B_M(i-1,j)
  &\mathrel{+}=
  p_dB_X(i,j),
  \label{eq:supp-align-reverse-update-x-open}
  \\
  B_X(i-1,j)
  &\mathrel{+}=
  p_eB_X(i,j),
  \label{eq:supp-align-reverse-update-x-ext}
\end{align}
and for each \(Y\)-cell with \(j\ge1\),
\begin{align}
  B_M(i,j-1)
  &\mathrel{+}=
  p_dB_Y(i,j),
  \label{eq:supp-align-reverse-update-y-open}
  \\
  B_Y(i,j-1)
  &\mathrel{+}=
  p_eB_Y(i,j).
  \label{eq:supp-align-reverse-update-y-ext}
\end{align}
Processing cells in reverse topological order gives exactly
Eqs.~\eqref{eq:supp-align-backward-m}--\eqref{eq:supp-align-backward-y}.

\subsection{Posterior local events and expected feature counts}
\label{sec:supp-alignment-posteriors}

The normalized product of an item and its adjoint gives the posterior
probability that a sampled alignment path uses that item. In particular,
\begin{equation}
  P(x_i\sim y_j\mid x,y)
  =
  \frac{
    F_M(i,j)B_M(i,j)
  }{
    Z_{\mathrm{aln}}(x,y)
  }.
  \label{eq:supp-align-match-item-posterior}
\end{equation}
A local event must additionally include the corresponding predecessor item
and transition or emission factor. For the three possible predecessor
states of a match at \((i,j)\),
\begin{equation}
  P(U\to M\text{ at }(i,j)\mid x,y)
  =
  \frac{
    F_U(i-1,j-1)
    q(x_i,y_j)
    B_M(i,j)
  }{
    Z_{\mathrm{aln}}(x,y)
  },
  \qquad
  U\in\{M,X,Y\}.
  \label{eq:supp-align-match-local-event}
\end{equation}
Their sum is
Eq.~\eqref{eq:supp-align-match-item-posterior}.

The \(X\)-gap opening and extension posteriors are
\begin{align}
  P(M\to X\text{ at }(i,j)\mid x,y)
  &=
  \frac{
    F_M(i-1,j)p_dB_X(i,j)
  }{
    Z_{\mathrm{aln}}(x,y)
  },
  \label{eq:supp-align-x-open-posterior}
  \\
  P(X\to X\text{ at }(i,j)\mid x,y)
  &=
  \frac{
    F_X(i-1,j)p_eB_X(i,j)
  }{
    Z_{\mathrm{aln}}(x,y)
  }.
  \label{eq:supp-align-x-ext-posterior}
\end{align}
Similarly,
\begin{align}
  P(M\to Y\text{ at }(i,j)\mid x,y)
  &=
  \frac{
    F_M(i,j-1)p_dB_Y(i,j)
  }{
    Z_{\mathrm{aln}}(x,y)
  },
  \label{eq:supp-align-y-open-posterior}
  \\
  P(Y\to Y\text{ at }(i,j)\mid x,y)
  &=
  \frac{
    F_Y(i,j-1)p_eB_Y(i,j)
  }{
    Z_{\mathrm{aln}}(x,y)
  }.
  \label{eq:supp-align-y-ext-posterior}
\end{align}

Let \(N_M(a,b)\) count matched columns containing symbol pair \((a,b)\),
let \(N_{\mathrm{open}}\) count all \(M\to X\) and \(M\to Y\) transitions,
and let \(N_{\mathrm{ext}}\) count all \(X\to X\) and \(Y\to Y\)
transitions. Since
\begin{equation}
  \log q(a,b)
  =
  \frac{s(a,b)}{\tau},
  \qquad
  \log p_d
  =
  -\frac{d}{\tau},
  \qquad
  \log p_e
  =
  -\frac{e}{\tau},
  \label{eq:supp-align-log-factors}
\end{equation}
the exponential-family identities of Supplementary Section~S1 give
\begin{align}
  \frac{
    \partial\log Z_{\mathrm{aln}}
  }{
    \partial s(a,b)
  }
  &=
  \frac{1}{\tau}
  \mathbb{E}\!\left[
    N_M(a,b)
    \mid x,y
  \right],
  \label{eq:supp-align-score-count}
  \\
  \frac{
    \partial\log Z_{\mathrm{aln}}
  }{
    \partial d
  }
  &=
  -\frac{1}{\tau}
  \mathbb{E}\!\left[
    N_{\mathrm{open}}
    \mid x,y
  \right],
  \label{eq:supp-align-open-count}
  \\
  \frac{
    \partial\log Z_{\mathrm{aln}}
  }{
    \partial e
  }
  &=
  -\frac{1}{\tau}
  \mathbb{E}\!\left[
    N_{\mathrm{ext}}
    \mid x,y
  \right].
  \label{eq:supp-align-ext-count}
\end{align}
For example,
\begin{align}
  \mathbb{E}\!\left[
    N_{\mathrm{open}}
    \mid x,y
  \right]
  &=
  \sum_{i=1}^{L}
  \sum_{j=0}^{K}
  \frac{
    F_M(i-1,j)p_dB_X(i,j)
  }{
    Z_{\mathrm{aln}}
  }
  \nonumber\\
  &\quad+
  \sum_{i=0}^{L}
  \sum_{j=1}^{K}
  \frac{
    F_M(i,j-1)p_dB_Y(i,j)
  }{
    Z_{\mathrm{aln}}
  },
  \label{eq:supp-align-open-count-explicit}
\end{align}
where terms involving out-of-domain predecessor cells are omitted.
Summing the \(X\)- and \(Y\)-extension posteriors in
Eqs.~\eqref{eq:supp-align-x-ext-posterior} and
\eqref{eq:supp-align-y-ext-posterior} similarly gives the expectation of
\(N_{\mathrm{ext}}\).

\subsection{Exact one-site substitution and group affinity}
\label{sec:supp-alignment-one-site}

Consider a substitution \(x_i\to c\) under the score-only parameterization
of the main text. The only sequence-identity-dependent factors involving
\(x_i\) are
\begin{equation}
  \boldsymbol{\rho}^{M}_i
  =
  \left(
    \rho^{M}_{i,j}
  \right)_{j=1}^{K},
  \qquad
  \rho^{M}_{i,j}
  =
  q(x_i,y_j).
  \label{eq:supp-align-position-group}
\end{equation}
The coordinates in Eq.~\eqref{eq:supp-align-position-group} are treated as
position-specific factor copies even when two coordinates happen to have the
same numerical value because the corresponding symbols in \(y\) are equal.
Every alignment path either matches \(x_i\) to exactly one \(y_j\), in
which case it uses one coordinate of
\(\boldsymbol{\rho}^{M}_i\), or aligns \(x_i\) to a gap, in which case it
uses no coordinate from this group. Consequently,
\begin{equation}
  Z_{\mathrm{aln}}
  =
  C_i
  +
  \sum_{j=1}^{K}
  C_{i,j}
  \rho^{M}_{i,j},
  \label{eq:supp-align-affine-decomposition}
\end{equation}
where \(C_i\) collects paths in which \(x_i\) is aligned to a gap and the
coefficients \(C_{i,j}\) are independent of the whole group
\(\boldsymbol{\rho}^{M}_i\).

Differentiating
Eq.~\eqref{eq:supp-align-affine-decomposition} gives
\begin{equation}
  C_{i,j}
  =
  \frac{
    \partial Z_{\mathrm{aln}}
  }{
    \partial\rho^{M}_{i,j}
  }.
  \label{eq:supp-align-affine-coefficient-derivative}
\end{equation}
The local-event factorization gives the same coefficient explicitly:
\begin{equation}
  \frac{
    \partial Z_{\mathrm{aln}}
  }{
    \partial\rho^{M}_{i,j}
  }
  =
  \frac{
    F_M(i,j)B_M(i,j)
  }{
    q(x_i,y_j)
  }.
  \label{eq:supp-align-match-factor-derivative}
\end{equation}
Indeed, \(F_M(i,j)/q(x_i,y_j)\) contains only prefix paths before the
consumption of \(x_i\), while \(B_M(i,j)\) contains only suffix paths after
that consumption; neither depends on another coordinate of the same
position group.

Let
\begin{equation}
  \Delta\rho^{M}_{i,j}
  =
  q(c,y_j)-q(x_i,y_j).
  \label{eq:supp-align-factor-change}
\end{equation}
Group affinity therefore gives the exact finite replacement
\begin{align}
  &
  Z_{\mathrm{aln}}\!\left(
    x^{i\to c},
    y
  \right)
  -
  Z_{\mathrm{aln}}(x,y)
  \nonumber\\
  &\quad=
  \sum_{j=1}^{K}
  \frac{
    F_M(i,j)B_M(i,j)
  }{
    q(x_i,y_j)
  }
  \left[
    q(c,y_j)-q(x_i,y_j)
  \right].
  \label{eq:supp-align-exact-one-site}
\end{align}
After division by \(Z_{\mathrm{aln}}(x,y)\),
\begin{equation}
  \frac{
    Z_{\mathrm{aln}}\!\left(
      x^{i\to c},
      y
    \right)
    -
    Z_{\mathrm{aln}}(x,y)
  }{
    Z_{\mathrm{aln}}(x,y)
  }
  =
  \sum_{j=1}^{K}
  P(x_i\sim y_j\mid x,y)
  \left[
    \frac{q(c,y_j)}{q(x_i,y_j)}
    -
    1
  \right].
  \label{eq:supp-align-exact-one-site-posterior}
\end{equation}
No small-change approximation is used in
Eqs.~\eqref{eq:supp-align-exact-one-site}--\eqref{eq:supp-align-exact-one-site-posterior}.

The dependence in
Eq.~\eqref{eq:supp-align-affine-decomposition} is not generally
homogeneous because \(C_i\) need not vanish: paths in which \(x_i\) is
aligned to a gap do not use any match-factor coordinate from
\(\boldsymbol{\rho}^{M}_i\). This example therefore separates the two
conditions emphasized in the main text: group-wise affinity is sufficient
for exact one-site replacement, whereas degree-one homogeneity is a stronger
property.

\subsection{Parameterization dependence and pair-HMM variants}
\label{sec:supp-alignment-parameterizations}

The factor group in
Eq.~\eqref{eq:supp-align-position-group} is complete only for the
score-based model used in the main text, in which the gap factors \(p_d\)
and \(p_e\) do not depend on residue identity. In another parameterization,
a substitution \(x_i\to c\) may alter additional local factors, and all of
them must be included in the mutation-dependent position group.

For example, consider a three-state pair-HMM in which an \(M\)-state emits
a pair with factor \(e_M(x_i,y_j)\), whereas an \(X\)-state emits the
residue \(x_i\) with factor \(e_X(x_i)\). To analyze a sequence edit rather
than a change of a globally shared model parameter, introduce
position-specific copies
\begin{equation}
  \rho^{M}_{i,j}
  =
  e_M(x_i,y_j),
  \qquad
  \rho^{X}_{i}
  =
  e_X(x_i).
  \label{eq:supp-align-pairhmm-position-factors}
\end{equation}
The complete \(x_i\)-dependent group is then
\begin{equation}
  \boldsymbol{\rho}^{x}_i
  =
  \left(
    \rho^{X}_{i},
    \rho^{M}_{i,1},
    \ldots,
    \rho^{M}_{i,K}
  \right).
  \label{eq:supp-align-expanded-position-group}
\end{equation}
Every complete alignment path consumes \(x_i\) exactly once, either in an
\(X\)-event or in one \(M\)-event. Hence, for this expanded group and this
parameterization, the partition function is homogeneous of degree one:
\begin{equation}
  Z_{\mathrm{aln}}\!\left(
    \ldots,
    \lambda\boldsymbol{\rho}^{x}_i,
    \ldots
  \right)
  =
  \lambda
  Z_{\mathrm{aln}}\!\left(
    \ldots,
    \boldsymbol{\rho}^{x}_i,
    \ldots
  \right).
  \label{eq:supp-align-pairhmm-homogeneity}
\end{equation}
The corresponding finite substitution remains one exact adjoint
contraction, but it contains both the match-emission changes and the
\(X\)-emission change.

This comparison shows that ``alignment is non-homogeneous'' is not a
model-invariant statement. The invariant requirement is to identify the
complete set of local factors changed by the sequence edit and determine
whether the partition function is affine in that group. Tying several
position-specific copies to one shared model parameter is useful for
parameter learning, but the sequence-to-factor analysis conceptually
unties them by position before asking how a particular observed symbol is
replaced.

\subsection{Multiple substitutions and mixed alignment effects}
\label{sec:supp-alignment-multisite}

Now substitute two distinct positions \(i<k\) of \(x\),
\begin{equation}
  x_i\to c,
  \qquad
  x_k\to d.
  \label{eq:supp-align-two-site-edit}
\end{equation}
Under the score-only model, define
\begin{equation}
  \Delta\rho^{M}_{i,j}
  =
  q(c,y_j)-q(x_i,y_j),
  \qquad
  \Delta\rho^{M}_{k,\ell}
  =
  q(d,y_\ell)-q(x_k,y_\ell).
  \label{eq:supp-align-two-site-factor-changes}
\end{equation}
The alignment partition function is affine in each position group
separately and hence multi-affine across the groups. Therefore
\begin{align}
  \Delta_{i,k}Z_{\mathrm{aln}}
  &=
  \sum_{j=1}^{K}
  \frac{
    \partial Z_{\mathrm{aln}}
  }{
    \partial\rho^{M}_{i,j}
  }
  \Delta\rho^{M}_{i,j}
  \nonumber\\
  &\quad+
  \sum_{\ell=1}^{K}
  \frac{
    \partial Z_{\mathrm{aln}}
  }{
    \partial\rho^{M}_{k,\ell}
  }
  \Delta\rho^{M}_{k,\ell}
  \nonumber\\
  &\quad+
  \sum_{j=1}^{K}
  \sum_{\ell=1}^{K}
  \frac{
    \partial^2 Z_{\mathrm{aln}}
  }{
    \partial\rho^{M}_{i,j}
    \partial\rho^{M}_{k,\ell}
  }
  \Delta\rho^{M}_{i,j}
  \Delta\rho^{M}_{k,\ell}.
  \label{eq:supp-align-two-site-expansion}
\end{align}
By path monotonicity, the mixed derivative in
Eq.~\eqref{eq:supp-align-two-site-expansion} is zero unless \(j<\ell\).
For \(j<\ell\), the normalized mixed derivative has the joint-posterior
interpretation
\begin{equation}
  \frac{
    \rho^{M}_{i,j}
    \rho^{M}_{k,\ell}
  }{
    Z_{\mathrm{aln}}
  }
  \frac{
    \partial^2 Z_{\mathrm{aln}}
  }{
    \partial\rho^{M}_{i,j}
    \partial\rho^{M}_{k,\ell}
  }
  =
  P\!\left(
    x_i\sim y_j,\,
    x_k\sim y_\ell
    \mid x,y
  \right).
  \label{eq:supp-align-joint-match-posterior}
\end{equation}
Hence the mixed term is the finite interaction arising from alignment paths
that use both mutation-dependent factor groups. It is not, in general,
recoverable by adding the two independently evaluated one-site effects.

More generally, for a set \(P\) of edited positions, exact replacement
requires the mixed derivative associated with every nonempty subset of
\(P\), exactly as in the general multi-affine expansion of Supplementary
Section~S1. The alignment topology adds a simple structural restriction:
joint match terms vanish whenever the proposed matched positions violate
the monotone ordering of an alignment path.

The first derivatives required for all one-site substitutions are obtained
from one forward and one reverse pass. Higher-order mixed derivatives can be
computed by additional dynamic-programming contractions or by successive
background updates, but their number grows with the order of the edit.
Thus the exactness of the multi-affine expansion does not imply that
arbitrary high-order mutation scans are computationally inexpensive.


\section{RNA ensemble differentiation at a fixed sequence}
\label{sec:supp-rna-fixed-sequence}

This section gives the RNA specialization supporting Sections~3.2 and~3.3
of the main text. Throughout this section the observed RNA sequence
\(x=x_1\cdots x_L\) is fixed. The objects differentiated here are
dynamic-programming items, local Boltzmann factors, thermodynamic
parameters, and coupled-grammar parameters. Finite changes of the sequence
itself are deliberately excluded and are treated in Supplementary
Section~S4.

A full Turner-style implementation contains many model-specific states and
cases for dangling ends, terminal mismatches, multiloops, and related
boundary conventions. Rather than reproduce one software implementation, we
use a decomposition-complete McCaskill-type formulation: every local case of
the chosen implementation is represented as one decomposition of one
dynamic-programming item. The adjoint and posterior formulas below then
apply unchanged after any required state expansion.

\subsection{Fixed-sequence thermodynamic ensemble and McCaskill-type decompositions}
\label{sec:supp-rna-inside}

Let \(\Omega_L\) be a common set of pseudoknot-free secondary-structure
candidates on the \(L\) labelled positions. For a local structural
configuration \(\ell\), define
\begin{equation}
  \phi_x(\ell)
  =
  \begin{cases}
    \displaystyle
    \exp\!\left\{
      -\frac{E_x(\ell)}{RT}
    \right\},
    &
    \ell\text{ is compatible with }x,
    \\[8pt]
    0,
    &
    \text{otherwise},
  \end{cases}
  \label{eq:supp-rna-local-factor}
\end{equation}
where \(E_x(\ell)\) is a molar free-energy contribution, \(R\) is the gas
constant, and \(T\) is the absolute temperature. The partition function is
\begin{equation}
  Z_{\mathrm{RNA}}(x)
  =
  \sum_{\sigma\in\Omega_L}
  \prod_{\ell\in\sigma}
  \phi_x(\ell).
  \label{eq:supp-rna-partition}
\end{equation}
The sequence remains fixed in all differentiations below, so zero factors in
Eq.~\eqref{eq:supp-rna-local-factor} can simply be regarded as excluded
local cases. The common-universe convention becomes important only when
different sequences are compared in Supplementary Section~S4.

A McCaskill-type algorithm evaluates
Eq.~\eqref{eq:supp-rna-partition} with a finite collection of interval
states. Let \(\mathcal{T}\) denote the state types used by the chosen
implementation. Typical types include an exterior state, a paired state,
and one or more multiloop states. For an item
\begin{equation}
  u
  =
  (\tau,i,j),
  \qquad
  \tau\in\mathcal{T},
  \label{eq:supp-rna-item}
\end{equation}
write \(Z^\tau(i,j)\), or more compactly \(A_u\), for its inside partition
function. Let \(\Gamma(u)\) be the complete set of local decompositions that
create \(u\). A decomposition \(\gamma\in\Gamma(u)\) has local factor
\(\phi_\gamma(x)\) and child-item multiset
\(\operatorname{ch}(\gamma)\). Then
\begin{equation}
  A_u
  =
  \sum_{\gamma\in\Gamma(u)}
  T_\gamma,
  \qquad
  T_\gamma
  =
  \phi_\gamma(x)
  \prod_{v\in\operatorname{ch}(\gamma)}
  A_v.
  \label{eq:supp-rna-general-inside}
\end{equation}
The root exterior item has value \(Z_{\mathrm{RNA}}(x)\).

For the paired state, write
\begin{equation}
  Z^b(i,j)
  =
  A_{(b,i,j)}.
  \label{eq:supp-rna-paired-state}
\end{equation}
Its decomposition set contains the usual topological classes:
a hairpin closed by \((i,j)\), an enclosed stack or interior/bulge loop
whose inner closing pair is \((k,\ell)\), and a multiloop closed by
\((i,j)\). A representative decomposition-level expression is
\begin{align}
  Z^b(i,j)
  &=
  \phi_{\mathrm{hp}}(i,j)
  \nonumber\\
  &\quad+
  \sum_{i<k<\ell<j}
  \phi_{\mathrm{int}}(i,j;k,\ell)
  Z^b(k,\ell)
  \nonumber\\
  &\quad+
  \phi_{\mathrm{ml}}(i,j)
  \mathcal{M}(i+1,j-1),
  \label{eq:supp-rna-paired-representative}
\end{align}
where \(\phi_{\mathrm{int}}\) includes stacking as the zero-unpaired
interior-loop case, and \(\mathcal{M}\) denotes the appropriate multiloop
inside expression of the selected state system. A common implementation
represents \(\mathcal{M}\) through states such as \(Z^m\) and \(Z^{m1}\);
other implementations split these states further to encode dangling-end or
boundary-context conventions.

The exterior and multiloop states likewise decompose only through
pseudoknot-free operations: an unpaired extension, insertion of a paired
branch, or concatenation of compatible subintervals. Hence every concrete
McCaskill or equivalent unambiguous grammar implementation can be written in
the form of Eq.~\eqref{eq:supp-rna-general-inside}. The precise energy model
changes the state set and local factors but not the adjoint identities below
\citep{McCaskill1990}.

\subsection{Outside quantities as reverse-mode adjoints}
\label{sec:supp-rna-outside}

For every RNA dynamic-programming item \(u\), define its outside adjoint by
\begin{equation}
  B_u
  =
  \frac{\partial Z_{\mathrm{RNA}}}{\partial A_u}.
  \label{eq:supp-rna-general-outside}
\end{equation}
For the paired state we use the conventional notation
\begin{equation}
  W^b(i,j)
  =
  \frac{
    \partial Z_{\mathrm{RNA}}
  }{
    \partial Z^b(i,j)
  }.
  \label{eq:supp-rna-paired-outside}
\end{equation}

The reverse update corresponding to one decomposition
\(\gamma\in\Gamma(u)\) is obtained by differentiating its local term
\(T_\gamma\). For every selected occurrence of child item \(v\) in
\(\operatorname{ch}(\gamma)\),
\begin{equation}
  B_v
  \mathrel{+}=
  B_u
  \phi_\gamma(x)
  \prod_{
    v'\in
    \operatorname{ch}(\gamma)
    \setminus
    \{v\}_{\mathrm{occ}}
  }
  A_{v'}.
  \label{eq:supp-rna-reverse-update}
\end{equation}
The root adjoint is initialized to one, all other adjoints to zero, and the
updates are processed in reverse topological order. Equation
\eqref{eq:supp-rna-reverse-update} is the complete outside algorithm at the
decomposition level: each state-specific outside recursion is the sum of
these updates over all parent decompositions in which the item occurs.

For example, the interior-loop term in
Eq.~\eqref{eq:supp-rna-paired-representative} gives
\begin{equation}
  W^b(k,\ell)
  \mathrel{+}=
  W^b(i,j)
  \phi_{\mathrm{int}}(i,j;k,\ell),
  \label{eq:supp-rna-interior-outside-update}
\end{equation}
for every allowed \(i<k<\ell<j\). The multiloop-closing term gives the
corresponding update into the root multiloop item,
\begin{equation}
  W^{\mathcal{M}}(i+1,j-1)
  \mathrel{+}=
  W^b(i,j)
  \phi_{\mathrm{ml}}(i,j),
  \label{eq:supp-rna-multiloop-outside-update}
\end{equation}
with further reverse propagation through the internal multiloop states
according to Eq.~\eqref{eq:supp-rna-reverse-update}. If an implementation
splits a local energy into additional states, the same chain rule simply
introduces additional reverse updates.

This formulation is useful because the outside quantity is defined by the
computational graph rather than by a separate probabilistic construction.
The biological meaning of \(W^b(i,j)\) is the total compatible weight of all
contexts outside the paired substructure \((i,j)\), while its mathematical
definition remains the adjoint in
Eq.~\eqref{eq:supp-rna-paired-outside}.

\subsection{Base-pair and motif-level posterior identities}
\label{sec:supp-rna-posteriors}

Let \(I_{ij}(\sigma)\) be the indicator that structure \(\sigma\) contains
base pair \((i,j)\). In an unambiguous McCaskill-type decomposition, each
structure containing \((i,j)\) uses the paired item \(Z^b(i,j)\) exactly
once. The general item-use identity of Supplementary Section~S1 therefore
gives
\begin{equation}
  P(i,j\mid x)
  =
  \mathbb{E}\!\left[
    I_{ij}
    \mid x
  \right]
  =
  \frac{
    Z^b(i,j)W^b(i,j)
  }{
    Z_{\mathrm{RNA}}(x)
  }.
  \label{eq:supp-rna-bpp}
\end{equation}
Thus the standard base-pairing probability is a normalized
inside--outside product.

Now fix one local decomposition
\(\gamma\in\Gamma(u)\). Let \(N_\gamma(\sigma)\) count its use in a
structure. Applying a marker to this single decomposition yields
\begin{equation}
  \mathbb{E}\!\left[
    N_\gamma
    \mid x
  \right]
  =
  \frac{
    B_uT_\gamma
  }{
    Z_{\mathrm{RNA}}(x)
  }
  =
  \frac{
    B_u
    \phi_\gamma(x)
    \displaystyle
    \prod_{v\in\operatorname{ch}(\gamma)}
    A_v
  }{
    Z_{\mathrm{RNA}}(x)
  }.
  \label{eq:supp-rna-local-event-posterior}
\end{equation}
For a decomposition tied to a fixed interval and split, this quantity is the
posterior probability of that particular loop or decomposition event.

Motif-level expected counts are obtained by summing local-event posteriors.
Let \(\mathcal{G}_m\) be a collection of decompositions that define motif
or local structural feature \(m\), and let
\begin{equation}
  N_m(\sigma)
  =
  \sum_{\gamma\in\mathcal{G}_m}
  N_\gamma(\sigma).
  \label{eq:supp-rna-motif-count}
\end{equation}
Then
\begin{equation}
  \mathbb{E}\!\left[
    N_m
    \mid x
  \right]
  =
  \sum_{\gamma\in\mathcal{G}_m}
  \frac{
    B_{\operatorname{pa}(\gamma)}
    T_\gamma
  }{
    Z_{\mathrm{RNA}}(x)
  },
  \label{eq:supp-rna-motif-expectation}
\end{equation}
where \(\operatorname{pa}(\gamma)\) denotes the item created by
\(\gamma\). Examples include expected counts of selected hairpin classes,
stacking configurations, interior-loop classes, or other features that can
be expressed as additive increments on local decompositions.

Equation~\eqref{eq:supp-rna-motif-expectation} also states precisely which
features are directly accessible without changing the dynamic-programming
state space: the feature must decompose additively over existing local
events. A feature that depends on additional nonlocal memory generally
requires state augmentation before the same expected-count identity can be
applied.

\subsection{Thermodynamic parameters as an exponential family}
\label{sec:supp-rna-exponential-family}

For fixed sequence \(x\), suppose that the structure energy is expressed as
\begin{equation}
  E_x(\sigma)
  =
  \sum_{k=1}^{K_\theta}
  \theta_k
  f_k(\sigma,x),
  \label{eq:supp-rna-energy-feature-expansion}
\end{equation}
where \(f_k\) is the count or accumulated value of thermodynamic feature
\(k\). Define the natural coordinates
\begin{equation}
  \eta_k
  =
  -\frac{\theta_k}{RT}.
  \label{eq:supp-rna-natural-parameter}
\end{equation}
The Boltzmann distribution becomes
\begin{equation}
  P(\sigma\mid x;\boldsymbol{\eta})
  =
  \exp\!\left\{
    \sum_k
    \eta_k
    f_k(\sigma,x)
    -
    \Psi(\boldsymbol{\eta})
  \right\},
  \qquad
  \Psi(\boldsymbol{\eta})
  =
  \log Z_{\mathrm{RNA}}(x).
  \label{eq:supp-rna-exponential-family}
\end{equation}
Therefore,
\begin{equation}
  \frac{
    \partial\Psi
  }{
    \partial\eta_k
  }
  =
  \mathbb{E}\!\left[
    f_k
    \mid x
  \right],
  \label{eq:supp-rna-gradient-expectation}
\end{equation}
or, in the original energy parameters,
\begin{equation}
  \frac{
    \partial\log Z_{\mathrm{RNA}}
  }{
    \partial\theta_k
  }
  =
  -\frac{1}{RT}
  \mathbb{E}\!\left[
    f_k
    \mid x
  \right].
  \label{eq:supp-rna-energy-parameter-gradient}
\end{equation}
This is the shared-feature version of the local-event identity in
Eq.~\eqref{eq:supp-rna-local-event-posterior}: if one parameter appears in
many loop decompositions, reverse differentiation accumulates all
corresponding local-event contributions.

Second derivatives give the covariance structure,
\begin{equation}
  \frac{
    \partial^2\Psi
  }{
    \partial\eta_k
    \partial\eta_\ell
  }
  =
  \operatorname{Cov}\!\left(
    f_k,
    f_\ell
    \mid x
  \right),
  \label{eq:supp-rna-hessian-covariance}
\end{equation}
and hence
\begin{equation}
  \frac{
    \partial^2\log Z_{\mathrm{RNA}}
  }{
    \partial\theta_k
    \partial\theta_\ell
  }
  =
  \frac{1}{(RT)^2}
  \operatorname{Cov}\!\left(
    f_k,
    f_\ell
    \mid x
  \right).
  \label{eq:supp-rna-energy-hessian}
\end{equation}
In natural coordinates the covariance matrix is the Fisher information
matrix of the exponential family, possibly singular when the feature
representation is nonminimal \citep{WainwrightJordan2008}. It is positive
semidefinite, so \(\Psi(\boldsymbol{\eta})\) is convex in the natural
parameters. This convexity concerns parameter space at fixed \(x\); it does
not imply convexity of an optimization problem over discrete RNA sequences.

\subsection{Sensitivities of ensemble observables}
\label{sec:supp-rna-observable-sensitivity}

The same exponential-family identity gives parameter sensitivities of
ensemble observables. Let \(g(\sigma,x)\) be an observable with no explicit
dependence on \(\eta_k\). Then
\begin{equation}
  \frac{
    \partial
    \mathbb{E}[g\mid x]
  }{
    \partial\eta_k
  }
  =
  \operatorname{Cov}\!\left(
    g,
    f_k
    \mid x
  \right).
  \label{eq:supp-rna-observable-covariance}
\end{equation}
Taking \(g=I_{ij}\) gives the sensitivity of a base-pairing probability:
\begin{equation}
  \frac{
    \partial P(i,j\mid x)
  }{
    \partial\eta_k
  }
  =
  \operatorname{Cov}\!\left(
    I_{ij},
    f_k
    \mid x
  \right).
  \label{eq:supp-rna-bpp-sensitivity}
\end{equation}
Taking \(g=f_\ell\) recovers
Eq.~\eqref{eq:supp-rna-hessian-covariance}.

If the observable itself depends explicitly on the parameter, the ordinary
product rule adds the direct term:
\begin{equation}
  \frac{
    \partial
    \mathbb{E}[g]
  }{
    \partial\eta_k
  }
  =
  \mathbb{E}\!\left[
    \frac{\partial g}{\partial\eta_k}
  \right]
  +
  \operatorname{Cov}\!\left(
    g,
    f_k
  \right).
  \label{eq:supp-rna-observable-explicit-dependence}
\end{equation}
These formulas distinguish parameter sensitivity at a fixed sequence from
the finite sequence-replacement problem considered in Supplementary
Section~S4.

\subsection{Expected-count learning in the RNAelem coupled grammar}
\label{sec:supp-rna-rnaelem}

RNAelem gives a concrete example in which the same fixed-sequence
inside--outside calculus is used for learning a sequence--structure model
\citep{MiyakeEtAl2024}. Let \(\varphi\) be a parse tree of the Profile CFG,
let \(\sigma\) be a parse tree of the RNA energy CFG, and let
\(\tau=\sigma\otimes\varphi\) be a compatible parse tree of the Coupled CFG.
For
\begin{equation}
  \nu
  =
  \left(
    \mu,
    \boldsymbol{\theta},
    \lambda
  \right),
  \label{eq:supp-rnaelem-parameters}
\end{equation}
write the unnormalized coupled weight in the form
\begin{equation}
  w_\nu(\tau;x)
  =
  \widetilde P_{\mathrm{profile}}\!\left(
    \varphi
    \mid
    \mu,\boldsymbol{\theta},x
  \right)
  \left[
    \widetilde P_{\mathrm{energy}}\!\left(
      \sigma
      \mid x
    \right)
  \right]^\lambda.
  \label{eq:supp-rnaelem-weight}
\end{equation}
Let \(\Phi(x)\) be the complete compatible parse-tree set and
\(\Phi_y(x)\subseteq\Phi(x)\) the constrained set corresponding to motif
presence or absence. Define
\begin{align}
  Z_\nu(x)
  &=
  \sum_{\tau\in\Phi(x)}
  w_\nu(\tau;x),
  \label{eq:supp-rnaelem-partition}
  \\
  Z_{\nu,y}(x)
  &=
  \sum_{\tau\in\Phi_y(x)}
  w_\nu(\tau;x).
  \label{eq:supp-rnaelem-constrained-partition}
\end{align}
Then
\begin{equation}
  P(y\mid\nu,x)
  =
  \frac{
    Z_{\nu,y}(x)
  }{
    Z_\nu(x)
  }.
  \label{eq:supp-rnaelem-class-probability}
\end{equation}

For a log-linear Profile-CFG parameter \(\theta_k\), let
\(N_k(\varphi,x)\) be its sufficient statistic. Differentiation of
Eq.~\eqref{eq:supp-rnaelem-class-probability} gives
\begin{align}
  \frac{
    \partial\log P(y\mid\nu,x)
  }{
    \partial\theta_k
  }
  &=
  \frac{
    \partial\log Z_{\nu,y}
  }{
    \partial\theta_k
  }
  -
  \frac{
    \partial\log Z_\nu
  }{
    \partial\theta_k
  }
  \nonumber\\
  &=
  \mathbb{E}_{\nu,y}\!\left[
    N_k
    \mid x
  \right]
  -
  \mathbb{E}_{\nu}\!\left[
    N_k
    \mid x
  \right].
  \label{eq:supp-rnaelem-profile-gradient}
\end{align}
Both expectations are computed by inside--outside algorithms on the Coupled
CFG; the first uses the constrained ensemble and the second the full
ensemble.

The scaling parameter \(\lambda\) has the same structure. Define
\begin{equation}
  S_{\mathrm{energy}}(\sigma,x)
  =
  \log
  \widetilde P_{\mathrm{energy}}(\sigma\mid x).
  \label{eq:supp-rnaelem-energy-statistic}
\end{equation}
Then
\begin{equation}
  \frac{
    \partial\log P(y\mid\nu,x)
  }{
    \partial\lambda
  }
  =
  \mathbb{E}_{\nu,y}\!\left[
    S_{\mathrm{energy}}
    \mid x
  \right]
  -
  \mathbb{E}_{\nu}\!\left[
    S_{\mathrm{energy}}
    \mid x
  \right].
  \label{eq:supp-rnaelem-lambda-gradient}
\end{equation}
For a labelled training set, these per-sequence gradients are summed and
combined with the derivative of the chosen regularization term.

RNAelem is therefore an application of the expected-count side of the
framework: model parameters vary while the observed sequence is fixed. It
does not require the finite sequence-to-factor replacement analysis used for
mutation scanning.

\subsection{Scope and transition to finite RNA sequence changes}
\label{sec:supp-rna-fixed-scope}

The identities in this section answer fixed-sequence questions:
base-pairing probabilities, local loop posteriors, expected structural
features, thermodynamic-parameter sensitivities, feature covariances, and
coupled-grammar learning. All are obtained from one sum--product computation
and its reverse adjoints.

A nucleotide substitution is a different operation. In a nearest-neighbor
energy model, one base can alter several primitive local factors that may
occur jointly, and it can change nucleotide contexts needed to evaluate
neighboring dynamic-programming states. Consequently, the mutant endpoint is
not generally obtained by replacing one affine position-specific factor
group in the original fixed-sequence computation.

Supplementary Section~S4 therefore starts from the inside, outside, and
local-event quantities defined here and asks a different question: how to
recombine them with mutation-consistent local factors and boundary contexts
to reconstruct the exact partition function of a mutant sequence.

\section{Finite RNA sequence changes and context-dependent recombination}
\label{sec:supp-rna-recombination}

Supplementary Section~S3 treated RNA ensemble differentiation with the
observed sequence held fixed. This section addresses the complementary
problem: the sequence itself is changed by a finite nucleotide
substitution. We first explain why an ordinary fixed-sequence
inside--local--outside contribution is not, in general, a complete mutation
formula. We then give a fully explicit one-site recombination for the
five-rule RNA-like SCFG used as the simplified RNA model in the numerical
tests. Finally, we introduce the boundary-context expansion underlying
Rchange, derive the ordered-background construction for two mutations, and
state precisely which exactness claims are verified here and which are
inherited from the published Rchange algorithm \citep{Kiryu2012}.

\subsection{From fixed-sequence adjoints to finite mutation recombination}
\label{sec:supp-rna-from-adjoint-to-recombination}

For a local RNA decomposition \(\gamma\) creating dynamic-programming item
\(u\), Supplementary Section~S3 gives the unnormalized contribution
\begin{equation}
  T_\gamma(x)
  =
  B_u(x)\,
  \phi_\gamma(x)
  \prod_{v\in\operatorname{ch}(\gamma)}
  A_v(x),
  \label{eq:supp-rna-original-local-contribution}
\end{equation}
where \(A_v\) are inside quantities and \(B_u\) is the outside adjoint.
After division by \(Z_{\mathrm{RNA}}(x)\), this is the posterior expected
occurrence of that local event.

If a hypothetical perturbation changed only the exposed factor
\(\phi_\gamma\), while leaving every reusable inside and outside quantity
unchanged, its finite contribution would be
\begin{equation}
  B_u(x)
  \left[
    \phi_\gamma\!\left(x^{p\to c}\right)
    -
    \phi_\gamma(x)
  \right]
  \prod_{v\in\operatorname{ch}(\gamma)}
  A_v(x).
  \label{eq:supp-rna-single-exposed-factor-change}
\end{equation}
Equation~\eqref{eq:supp-rna-single-exposed-factor-change} is exact for that
single exposed factor. A nucleotide substitution is more complicated.
One base can change several primitive thermodynamic factors that occur
jointly in one decomposition, and nearest-neighbor terms can depend on
nucleotides immediately outside an interval. The mutation can therefore
change both the local factor block and the boundary context under which an
otherwise reusable inside or outside quantity must be evaluated.

The finite-recombination problem is consequently to partition all latent
objects into nonoverlapping cases such that, in each case,

\begin{enumerate}
  \item the mutation-dependent primitive factors are exposed explicitly and
  replaced jointly;
  \item the remaining inside substructures are reusable after the appropriate
  mutant boundary contexts have been fixed; and
  \item the outside context is likewise selected consistently with the
  candidate nucleotide.
\end{enumerate}

This is the finite endpoint counterpart of the fixed-sequence adjoint
factorization. The simplest case is an RNA-like SCFG in which the nucleotide
at the mutated position is emitted exactly once and no neighboring-base
context is required.

\subsection{Explicit four-role recombination in a five-rule RNA-like SCFG}
\label{sec:supp-rna-five-rule-grammar}

Consider the one-nonterminal RNA-like grammar with rule types
\(P,L,R,B,E\):
\begin{align}
  P:&\quad S \longrightarrow a_{<}\,S\,a_{>},
  &
  L:&\quad S \longrightarrow a\,S,
  \nonumber\\
  R:&\quad S \longrightarrow S\,a,
  &
  B:&\quad S \longrightarrow S\,S,
  \qquad
  E:\quad S \longrightarrow \epsilon.
  \label{eq:supp-rna-five-rules}
\end{align}
Here \((a_{<},a_{>})\) is a paired emission, \(a\) is an unpaired
emission, and \(\epsilon\) is the empty string. We use the finite interval
dynamic program defined below: bifurcation splits contain two nonempty
spans, and the \(E\)-rule is used only as the empty-interval boundary
condition. Thus no cyclic insertion of empty bifurcation children is
included. The resulting derivation system is still ambiguous, but the
identities below are identities over its finite parse trees; distinct parse
trees are treated as distinct latent objects.

Let \(t_P,t_L,t_R,t_B,t_E\) be rule weights and let
\(e_P(a,b)\), \(e_L(a)\), and \(e_R(a)\) be emission factors. Define
\begin{equation}
  \tau_P(a,b)=t_Pe_P(a,b),
  \qquad
  \tau_L(a)=t_Le_L(a),
  \qquad
  \tau_R(a)=t_Re_R(a),
  \label{eq:supp-rna-local-rule-factors}
\end{equation}
with \(\tau_B=t_B\) and \(\tau_E=t_E\).

For \(x=x_1\cdots x_L\), let \(\alpha(i,j)\) be the inside weight for
deriving \(x_i\cdots x_j\). Empty intervals satisfy
\begin{equation}
  \alpha(i,i-1)=\tau_E.
  \label{eq:supp-rna-empty-inside}
\end{equation}
For \(i\le j\),
\begin{align}
  \alpha(i,j)
  &=
  \mathbf{1}\{i<j\}
  \tau_P(x_i,x_j)
  \alpha(i+1,j-1)
  +
  \tau_L(x_i)
  \alpha(i+1,j)
  \nonumber\\
  &\quad+
  \tau_R(x_j)
  \alpha(i,j-1)
  +
  \tau_B
  \sum_{k=i}^{j-1}
  \alpha(i,k)
  \alpha(k+1,j).
  \label{eq:supp-rna-scfg-inside}
\end{align}
The partition function over parse trees is
\begin{equation}
  Z_0(x)=\alpha(1,L).
  \label{eq:supp-rna-scfg-partition}
\end{equation}

Define the outside adjoint
\begin{equation}
  \beta(i,j)
  =
  \frac{\partial Z_0(x)}{\partial\alpha(i,j)},
  \label{eq:supp-rna-scfg-outside-definition}
\end{equation}
with \(\beta(1,L)=1\). Reverse accumulation through
Eq.~\eqref{eq:supp-rna-scfg-inside} gives
\begin{align}
  \beta(i+1,j-1)
  &\mathrel{+}=
  \mathbf{1}\{i<j\}
  \beta(i,j)\tau_P(x_i,x_j),
  \\
  \beta(i+1,j)
  &\mathrel{+}=
  \beta(i,j)\tau_L(x_i),
  \\
  \beta(i,j-1)
  &\mathrel{+}=
  \beta(i,j)\tau_R(x_j),
\end{align}
and, for every \(k=i,\ldots,j-1\),
\begin{align}
  \beta(i,k)
  &\mathrel{+}=
  \beta(i,j)\tau_B\alpha(k+1,j),
  \\
  \beta(k+1,j)
  &\mathrel{+}=
  \beta(i,j)\tau_B\alpha(i,k).
\end{align}
Updates to empty intervals can be omitted when \(\tau_E\) is fixed.

Fix position \(p\) and replace \(x_p\) by candidate nucleotide \(c\).
Every parse tree emits position \(p\) exactly once, through one of four
roles: the left nucleotide of a paired rule, the right nucleotide of a
paired rule, the left unpaired rule \(L\), or the right unpaired rule \(R\).
Define
\begin{align}
  \mathcal{R}^{\mathrm{PL}}_{p,x}(c)
  &:=
  \sum_{j=p+1}^{L}
  \beta(p,j)\tau_P(c,x_j)\alpha(p+1,j-1),
  \label{eq:supp-rna-role-paired-left}
  \\
  \mathcal{R}^{\mathrm{PR}}_{p,x}(c)
  &:=
  \sum_{i=1}^{p-1}
  \beta(i,p)\tau_P(x_i,c)\alpha(i+1,p-1),
  \label{eq:supp-rna-role-paired-right}
  \\
  \mathcal{R}^{\mathrm{UL}}_{p,x}(c)
  &:=
  \sum_{j=p}^{L}
  \beta(p,j)\tau_L(c)\alpha(p+1,j),
  \label{eq:supp-rna-role-unpaired-left}
  \\
  \mathcal{R}^{\mathrm{UR}}_{p,x}(c)
  &:=
  \sum_{i=1}^{p}
  \beta(i,p)\tau_R(c)\alpha(i,p-1).
  \label{eq:supp-rna-role-unpaired-right}
\end{align}
The explicit recombination operator is
\begin{align}
  \mathcal{R}_{p,x}(c)
  &:=
  \mathcal{R}^{\mathrm{PL}}_{p,x}(c)
  +
  \mathcal{R}^{\mathrm{PR}}_{p,x}(c)
  \nonumber\\
  &\quad+
  \mathcal{R}^{\mathrm{UL}}_{p,x}(c)
  +
  \mathcal{R}^{\mathrm{UR}}_{p,x}(c).
  \label{eq:supp-rna-explicit-four-role-recombination}
\end{align}

All inside and outside quantities in
Eq.~\eqref{eq:supp-rna-explicit-four-role-recombination} are computed in
the reference sequence. Once the unique rule occurrence emitting position
\(p\) is cut, the remaining inside and outside parse-tree fragments do not
contain the emission factor at \(p\), so they can be reused for every
candidate \(c\).

Because the four role classes are mutually exclusive and exhaustive,
\begin{equation}
  Z_0\!\left(x^{p\to c}\right)
  =
  \mathcal{R}_{p,x}(c),
  \qquad
  Z_0(x)
  =
  \mathcal{R}_{p,x}(x_p),
  \label{eq:supp-rna-four-role-endpoints}
\end{equation}
and
\begin{equation}
  \Delta_pZ_0
  =
  \mathcal{R}_{p,x}(c)
  -
  \mathcal{R}_{p,x}(x_p).
  \label{eq:supp-rna-four-role-exact-difference}
\end{equation}
At \(c=x_p\),
\begin{equation}
  \frac{
    \mathcal{R}^{\mathrm{PL}}_{p,x}(x_p)
    +\mathcal{R}^{\mathrm{PR}}_{p,x}(x_p)
    +\mathcal{R}^{\mathrm{UL}}_{p,x}(x_p)
    +\mathcal{R}^{\mathrm{UR}}_{p,x}(x_p)
  }{
    Z_0(x)
  }
  =1.
  \label{eq:supp-rna-role-posterior-normalization}
\end{equation}
Each term is a sum of fixed-sequence local-event posteriors of the type
derived in Supplementary Section~S3. This context-independent grammar is an
exactly-once, group-affine special case of Supplementary Section~S1.

\subsection{Boundary-context expansion and the Rchange construction}
\label{sec:supp-rna-boundary-contexts}

In nearest-neighbor RNA energy models, some local energies depend on bases
immediately outside the dynamic-programming interval. Rchange handles this
by introducing context-indexed copies of the inside and outside quantities
\citep{Kiryu2012}. In the simple SCFG construction used to introduce
the method, a boundary pair \((a,b)\) is expanded to the seven
single-mutation contexts
\begin{equation}
  \mathcal{C}(a,b)
  =
  \{(a,b)\}
  \cup
  \{(c,b):c\in\Sigma\setminus\{a\}\}
  \cup
  \{(a,c):c\in\Sigma\setminus\{b\}\}.
  \label{eq:supp-rna-seven-contexts}
\end{equation}
For a four-letter RNA alphabet,
\begin{equation}
  \left|\mathcal{C}(a,b)\right|=7.
  \label{eq:supp-rna-seven-context-size}
\end{equation}
Thus single-mutation context tracking enlarges the corresponding
inside--outside state family by a constant factor. In the Rfold energy-model
application of Rchange, dangling-end and terminal-mismatch dependencies do
not extend more than one nucleotide beyond either side of the relevant cell,
so the same finite-context principle applies.

To write the construction independently of a particular state table, let
\(\mathcal{T}_{\mathrm{Rch}}\) be the state types of the chosen RNA
partition-function algorithm. For state type \(\tau\) and item \(u\), write
\begin{equation}
  Z^\tau_u
  \!\left(
    \kappa^{\mathrm{in}};x_{-p}
  \right),
  \qquad
  W^\tau_u
  \!\left(
    \kappa^{\mathrm{out}};x_{-p}
  \right)
  \label{eq:supp-rna-context-indexed-items}
\end{equation}
for inside and outside quantities indexed by the boundary context required
by that state. The notation \(x_{-p}\) means that all unchanged sequence
symbols are fixed, while the nucleotide at \(p\) is supplied through the
context arguments and exposed local factors.

Let \(\mathcal{U}^{\tau}_p\) be the items of type \(\tau\) whose relevant
decomposition exposes position \(p\), or exposes a mutation-dependent
boundary context containing \(p\). For each such item \(u\), let
\(\Gamma^\tau_{p,u}\) be the compatible recombination cases. A case
\(\gamma\in\Gamma^\tau_{p,u}\) specifies its child items, their state
types, the outside and inside boundary contexts induced by candidate
nucleotide \(c\), and the joint product
\(\Phi_\gamma(c;\kappa_\gamma(c))\) of all mutation-dependent primitive
factors exposed in that case.

The primitive factors included in \(\Phi_\gamma\) are removed from the
reusable inside and outside pieces and inserted jointly. The general
Rchange-type recombination is
\begin{align}
  R^{\mathrm{Rch}}_{p,x}(c)
  &=
  \sum_{\tau\in\mathcal{T}_{\mathrm{Rch}}}
  \sum_{u\in\mathcal{U}^{\tau}_p}
  \sum_{\gamma\in\Gamma^\tau_{p,u}}
  W^\tau_u
  \!\left(
    \kappa^{\mathrm{out}}_\gamma(c);x_{-p}
  \right)
  \Phi_\gamma
  \!\left(
    c;\kappa_\gamma(c)
  \right)
  \nonumber\\
  &\qquad\qquad\times
  \prod_{v\in\operatorname{ch}(\gamma)}
  Z^{\tau_{\gamma,v}}_{u_{\gamma,v}}
  \!\left(
    \kappa^{\mathrm{in}}_{\gamma,v}(c);x_{-p}
  \right).
  \label{eq:supp-rna-general-rchange-recombination}
\end{align}
The sums in
Eq.~\eqref{eq:supp-rna-general-rchange-recombination} stand for the actual
state types, intervals, split points, loop classes, and boundary
configurations used by the implementation. Its common algebraic structure is
an outside context, a jointly replaced mutant local block, and compatible
inside children.

When the recombination cases are mutually exclusive and exhaustive for the
underlying algorithm,
\begin{equation}
  Z_{\mathrm{RNA}}\!\left(x^{p\to c}\right)
  =
  R^{\mathrm{Rch}}_{p,x}(c),
  \qquad
  Z_{\mathrm{RNA}}(x)
  =
  R^{\mathrm{Rch}}_{p,x}(x_p),
  \label{eq:supp-rna-rchange-endpoints}
\end{equation}
and therefore
\begin{equation}
  \Delta_p Z_{\mathrm{RNA}}
  =
  R^{\mathrm{Rch}}_{p,x}(c)
  -
  R^{\mathrm{Rch}}_{p,x}(x_p).
  \label{eq:supp-rna-rchange-one-site-difference}
\end{equation}

Equation~\eqref{eq:supp-rna-general-rchange-recombination} is a schematic
representation of the common Rchange construction, not a replacement for
the complete Rfold/Turner state-specific case table. The published
algorithmic exactness relies on a complete and nonredundant enumeration of
those cases. The full Rchange energy-model implementation uses the
unambiguous Rfold dynamic program; Rfold is not itself an SCFG, although its
inside--outside calculations have the same sum--product and reverse-adjoint
structure \citep{Kiryu2012}.

The connection with Supplementary Section~S1 is immediate in the
context-independent limit. If the boundary contexts do not depend on
\(c\), and every recombination case exposes one coordinate from one affine
mutation-dependent factor group, then the inside and outside pieces in
Eq.~\eqref{eq:supp-rna-general-rchange-recombination} become fixed
derivative coefficients. Equation~\eqref{eq:supp-rna-rchange-one-site-difference}
then reduces to the exact one-group affine replacement formula. The
four-role RNA-like SCFG of Supplementary Section~S4.2 is one explicit
instance of this limit.

\subsection{Two mutations as ordered background recombination}
\label{sec:supp-rna-double-recombination}

Let \(p\neq q\), with substitutions
\begin{equation}
  x_p\to c,
  \qquad
  x_q\to d.
  \label{eq:supp-rna-two-mutation-definition}
\end{equation}
Define the \(q\)-mutant background
\begin{equation}
  x'
  =
  x^{q\to d}.
  \label{eq:supp-rna-q-mutant-background}
\end{equation}
Applying the one-site recombination at \(p\) in this background gives the
double-mutant endpoint
\begin{equation}
  Z_{\mathrm{RNA}}\!\left(
    x^{p\to c,q\to d}
  \right)
  =
  R^{\mathrm{Rch}}_{p,x'}(c).
  \label{eq:supp-rna-double-mutant-endpoint}
\end{equation}
The inside and outside quantities used in
\(R^{\mathrm{Rch}}_{p,x'}\) must be appropriate to the \(q\)-mutant
background. They may be recomputed or obtained by updating the cells whose
decompositions depend on the first mutation.

The mixed finite difference is
\begin{align}
  \Delta_p\Delta_q Z_{\mathrm{RNA}}
  &=
  Z_{\mathrm{RNA}}\!\left(
    x^{p\to c,q\to d}
  \right)
  -
  Z_{\mathrm{RNA}}\!\left(
    x^{p\to c}
  \right)
  \nonumber\\
  &\quad-
  Z_{\mathrm{RNA}}\!\left(
    x^{q\to d}
  \right)
  +
  Z_{\mathrm{RNA}}(x).
  \label{eq:supp-rna-mixed-finite-difference}
\end{align}
Using the \(p\)-recombination operator in the two backgrounds,
\begin{align}
  \Delta_p\Delta_q Z_{\mathrm{RNA}}
  &=
  \left[
    R^{\mathrm{Rch}}_{p,x'}(c)
    -
    R^{\mathrm{Rch}}_{p,x'}(x_p)
  \right]
  \nonumber\\
  &\quad-
  \left[
    R^{\mathrm{Rch}}_{p,x}(c)
    -
    R^{\mathrm{Rch}}_{p,x}(x_p)
  \right].
  \label{eq:supp-rna-explicit-mixed-background}
\end{align}
Thus the mixed interaction is the change in the signed effect of the
\(p\)-mutation after the \(q\)-mutation has changed the background.

Applying the mutations in the opposite order gives
\begin{equation}
  Z_{\mathrm{RNA}}\!\left(
    x^{p\to c,q\to d}
  \right)
  =
  R^{\mathrm{Rch}}_{q,x^{p\to c}}(d).
  \label{eq:supp-rna-opposite-order-endpoint}
\end{equation}
The two expressions are different computational decompositions of the same
endpoint and agree up to numerical round-off. The total two-site change is
\begin{equation}
  \Delta_{p,q}Z_{\mathrm{RNA}}
  =
  \Delta_pZ_{\mathrm{RNA}}
  +
  \Delta_qZ_{\mathrm{RNA}}
  +
  \Delta_p\Delta_qZ_{\mathrm{RNA}}.
  \label{eq:supp-rna-total-two-site-change}
\end{equation}

Two mechanisms can contribute to the mixed term. The two positions can
enter the same primitive local factor, or the first mutation can change the
boundary context in which the second is evaluated. These mechanisms do not
define a generally unique additive decomposition of
\(\Delta_p\Delta_qZ_{\mathrm{RNA}}\).

\subsection{Computational scope and relation to the numerical verification}
\label{sec:supp-rna-verification-scope}

For the Rfold implementation with maximum base-pair span \(W\), the base
inside--outside computation has time complexity
\begin{equation}
  O(LW^2).
  \label{eq:supp-rna-rchange-base-complexity}
\end{equation}
The published exhaustive single-mutant calculation also has
\(O(LW^2)\) time complexity for the fixed four-letter RNA alphabet
\citep{Kiryu2012}. Over the complete \(O(L)\) set of one-site
candidates, this corresponds to the amortized \(O(W^2)\) per-candidate cost
reported in the main text.

For double mutations, unrestricted exhaustive evaluation requires
\begin{equation}
  O(L^2W^2),
  \label{eq:supp-rna-rchange-double-complexity}
\end{equation}
and restricting the maximum separation of the two mutated positions to
\(D\) gives
\begin{equation}
  O(LW^2D).
  \label{eq:supp-rna-rchange-distance-complexity}
\end{equation}
These complexity statements are properties of the published Rchange/Rfold
construction, not universal bounds for arbitrary RNA energy models or
context-state parameterizations.

The RNA numerical tests reported in the main text use the
context-independent five-rule grammar of Supplementary Section~S4.2. They
check
Eq.~\eqref{eq:supp-rna-four-role-exact-difference} for all one-site
candidates, the self-consistency identity
\begin{equation}
  \mathcal{R}_{p,x}(x_p)
  =
  Z_0(x),
  \label{eq:supp-rna-numerical-self-consistency}
\end{equation}
and the ordered-background construction for two mutations. Machine-precision
agreement therefore verifies the implementation of the explicit four-role
SCFG recombination and its multisite extension.

Those experiments do not independently revalidate every context-indexed
Rfold case of the full Turner-style Rchange implementation. The two levels
should be kept distinct:
\begin{itemize}
  \item the five-rule RNA-like SCFG formulas above are explicit identities
  derived and numerically checked in this work;
  \item the full nearest-neighbor result is the algorithmic exactness of the
  published Rchange construction, summarized abstractly by
  Eq.~\eqref{eq:supp-rna-general-rchange-recombination}.
\end{itemize}

This distinction connects the adjoint quantities of Supplementary
Section~S3 to exact finite RNA mutation calculations without claiming a new
rederivation or independent validation of the complete Turner/Rfold case
enumeration.

\section{Numerical verification and reproducibility}
\label{sec:supp-numerical-verification}

This section documents the numerical checks summarized in Section~5.2 of the
main text. The calculations are implementation checks of identities derived
analytically in the main text and Supplementary Sections~S1--S4; they are not
used as evidence for exactness. For each model, a formula-based mutant value
or finite difference was compared with direct reconstruction of the mutant
sequence followed by a complete rerun of the corresponding dynamic program.

The complete implementation is provided as Supplementary Code
\texttt{verify\_numerics.py}. The fixed instances, random-parameter
generation rules, test enumeration, and reference numerical outputs are
recorded below so that the numerical claims can be reproduced independently
of the prose description.

\subsection{Verification protocol and residual definitions}
\label{sec:supp-numerical-protocol}

Four model classes were tested:

\begin{enumerate}
  \item a three-state HMM on a length-eight sequence over a four-letter
  alphabet;
  \item a three-state affine-gap alignment ensemble for sequences of lengths
  four and three;
  \item a three-nonterminal SCFG with binary and lexical productions on a
  length-five sequence over a two-letter alphabet; and
  \item the five-rule RNA-like SCFG of Supplementary Section~S4.2 on a
  length-six RNA sequence.
\end{enumerate}

For a formula value \(Y_{\mathrm{formula}}\) and the corresponding value
\(Y_{\mathrm{direct}}\) obtained by a complete rerun, we report
\begin{equation}
  r_{\mathrm{exact}}
  =
  \frac{
    \left|
      Y_{\mathrm{formula}}-Y_{\mathrm{direct}}
    \right|
  }{
    Z(x)
  }.
  \label{eq:supp-numerical-residual}
\end{equation}
For HMM, alignment, and ordinary SCFG one-site checks,
\(Y\) is the finite change \(\Delta Z\). For the RNA-like recombination
check, \(Y\) is the reconstructed mutant endpoint partition function.
The same normalization by the original partition function \(Z(x)\) is used
throughout.

For every fixed model instance, all positions and all alternative symbols
were enumerated. The HMM and RNA-like SCFG tests additionally enumerate all
ordered replacement-symbol combinations for every pair of distinct positions
with the position pair represented once as \(p<q\). For the RNA-like SCFG,
the self-consistency identity
\begin{equation}
  \mathcal{R}_{p,x}(x_p)=Z_0(x)
  \label{eq:supp-numerical-rna-self}
\end{equation}
and both orders of successive background recombination were also checked.

\subsection{Fixed test instances}
\label{sec:supp-numerical-fixed-instances}

\subsubsection{HMM}

The alphabet is encoded as
\begin{equation}
  0=A,\qquad 1=C,\qquad 2=G,\qquad 3=U,
\end{equation}
and the observed sequence is
\begin{equation}
  x=(0,1,2,3,1,0,2,1).
  \label{eq:supp-numerical-hmm-sequence}
\end{equation}
The transition matrix, emission matrix, and initial-state vector are
\begin{equation}
  A=
  \begin{pmatrix}
    0.5 & 0.3 & 0.2\\
    0.2 & 0.6 & 0.2\\
    0.3 & 0.3 & 0.4
  \end{pmatrix},
  \qquad
  E=
  \begin{pmatrix}
    0.4  & 0.3  & 0.2  & 0.1\\
    0.1  & 0.4  & 0.4  & 0.1\\
    0.25 & 0.25 & 0.25 & 0.25
  \end{pmatrix},
  \label{eq:supp-numerical-hmm-matrices}
\end{equation}
and
\begin{equation}
  \boldsymbol{\pi}
  =
  (0.5,0.3,0.2).
  \label{eq:supp-numerical-hmm-pi}
\end{equation}
The forward initialization is
\(f_s(1)=\pi_s e_s(x_1)\), and the terminal backward values are one.
Every one-site alternative is checked against the exact adjoint contraction.
For every pair \(p<q\), every pair of alternative symbols is checked against
the exact two-site expansion including the mixed term.

\subsubsection{Affine-gap alignment}

The same four-letter encoding is used. The fixed sequences are
\begin{equation}
  x=(0,2,1,3)=\mathrm{AGCU},
  \qquad
  y=(0,3,1)=\mathrm{AUC}.
  \label{eq:supp-numerical-alignment-sequences}
\end{equation}
To avoid an ad hoc nucleotide scoring scheme, the fixed instance uses
the nucleotide substitution scores and affine-gap penalty magnitudes of
EMBOSS Needle \citep{RiceEtAl2000}. Needle uses the EDNAFULL matrix
(NCBI NUC4.4) by default for nucleic-acid sequences, with a gap-opening
penalty of \(10.0\) and a gap-extension penalty of \(0.5\).
Restricted to the unambiguous RNA alphabet \(\{A,C,G,U\}\), with \(U\)
identified with \(T\), EDNAFULL gives
\begin{equation}
  s(a,b)
  =
  \begin{cases}
    5, & a=b,\\
    -4, & a\neq b.
  \end{cases}
  \label{eq:supp-numerical-alignment-score}
\end{equation}
The affine-gap penalties are therefore
\begin{equation}
  d=10.0,
  \qquad
  e=0.5.
  \label{eq:supp-numerical-alignment-gap-parameters}
\end{equation}
Our alignment ensemble converts these conventional scores into positive
factors through
\begin{equation}
  q(a,b)=\exp\!\left(\frac{s(a,b)}{\tau}\right),
  \qquad
  p_d=\exp\!\left(-\frac{d}{\tau}\right),
  \qquad
  p_e=\exp\!\left(-\frac{e}{\tau}\right).
  \label{eq:supp-numerical-alignment-boltzmann-factors}
\end{equation}
We set
\begin{equation}
  \tau=5.0
  \label{eq:supp-numerical-alignment-scale}
\end{equation}
for the numerical audit. The scale \(\tau\) is a formal parameter of the
Boltzmann ensemble used in this paper and is not part of the EMBOSS scoring
convention; the value \(5.0\) keeps the exponentiated factors in a
numerically moderate range while preserving the chosen EDNAFULL/EMBOSS score and gap
ratios.

We retain the global-alignment boundary conditions of Supplementary
Section~S2, including its treatment of terminal gaps; only the substitution
matrix and affine-gap penalty magnitudes are taken from the EMBOSS defaults.
Thus this test is not intended to reproduce the complete command-line
behavior of Needle. Only symbols of \(x\) are mutated; \(y\) is held
fixed.

\subsubsection{Binary--lexical SCFG}

The nonterminals are indexed as
\begin{equation}
  0=S,\qquad 1=A,\qquad 2=B,
\end{equation}
and the terminal alphabet is \(\{0,1\}\). The observed sequence is
\begin{equation}
  x=(0,1,0,1,1).
  \label{eq:supp-numerical-scfg-sequence}
\end{equation}
The nonzero binary-rule weights are
\begin{align}
  t_{S\to AB}&=0.4,
  &
  t_{S\to BA}&=0.3,
  &
  t_{S\to AA}&=0.3,
  \nonumber\\
  t_{A\to BB}&=0.5,
  &
  t_{A\to SB}&=0.2,
  &
  t_{B\to AS}&=0.4,
  &
  t_{B\to BB}&=0.1.
  \label{eq:supp-numerical-scfg-rules}
\end{align}
All other binary-rule weights are zero. The lexical factors are
\begin{equation}
  E=
  \begin{pmatrix}
    0.5 & 0.5\\
    0.7 & 0.3\\
    0.2 & 0.8
  \end{pmatrix},
  \label{eq:supp-numerical-scfg-emissions}
\end{equation}
where row \(v\) gives \(e_v(0)\) and \(e_v(1)\).
The root is nonterminal \(S\). Every one-site replacement is compared with
the outside-coefficient formula using \(\beta(p,p,v)\).

\subsubsection{Five-rule RNA-like SCFG}

The sequence is
\begin{equation}
  x=(A,G,C,U,G,A).
  \label{eq:supp-numerical-rna-sequence}
\end{equation}
For the five rule types \(P,L,R,B,E\) of Supplementary Section~S4.2, the
fixed rule weights are
\begin{equation}
  t_P=0.30,\qquad
  t_L=0.25,\qquad
  t_R=0.20,\qquad
  t_B=0.15,\qquad
  t_E=1.0.
  \label{eq:supp-numerical-rna-rule-weights}
\end{equation}
The left- and right-unpaired emission factors, in the order
\((A,C,G,U)\), are
\begin{equation}
  e_L=(0.30,0.20,0.25,0.25),
  \qquad
  e_R=(0.22,0.28,0.24,0.26).
  \label{eq:supp-numerical-rna-unpaired}
\end{equation}
The paired-emission factors are
\begin{equation}
  e_P(A,U)=e_P(U,A)=1.0,
  \qquad
  e_P(G,C)=e_P(C,G)=1.6,
  \label{eq:supp-numerical-rna-pair-strong}
\end{equation}
\begin{equation}
  e_P(G,U)=e_P(U,G)=0.6,
  \label{eq:supp-numerical-rna-pair-wobble}
\end{equation}
with \(e_P(a,b)=0\) for all other ordered pairs.
The local factors used by the recursion are
\(t_Pe_P(a,b)\), \(t_Le_L(a)\), \(t_Re_R(a)\), \(t_B\), and \(t_E\).

The one-site test evaluates the four-role recombination operator
\(\mathcal{R}_{p,x}(c)\) for every candidate nucleotide. The two-site test
first constructs one single-mutant background and then evaluates the second
mutation by the same one-site recombination, repeating the calculation in
the opposite mutation order.

\subsection{Fixed-instance numerical results}
\label{sec:supp-numerical-fixed-results}

The partition functions and maximum normalized residuals obtained in the
reference audit are shown in Table~\ref{tab:supp-numerical-fixed-results}.
The residuals are at double-precision round-off.

\begin{table}[t]
\centering
\small
\caption{Fixed-instance numerical checks. Residuals are normalized by the
original partition function as in
Eq.~\eqref{eq:supp-numerical-residual}.}
\label{tab:supp-numerical-fixed-results}
\begin{tabular}{
  >{\raggedright\arraybackslash}p{0.18\textwidth}
  >{\raggedright\arraybackslash}p{0.17\textwidth}
  >{\raggedright\arraybackslash}p{0.31\textwidth}
  >{\raggedright\arraybackslash}p{0.17\textwidth}
}
\hline
Model & \(Z(x)\) & Checked identity & Maximum residual\\
\hline
HMM
& \(2.563540\times10^{-5}\)
& one-site replacement
& \(2.64\times10^{-16}\)\\
HMM
& \(2.563540\times10^{-5}\)
& two-site mixed expansion
& \(1.59\times10^{-15}\)\\
Affine-gap alignment
& \(8.044078\times10^{-1}\)
& one-site replacement
& \(1.38\times10^{-16}\)\\
SCFG
& \(1.396501\times10^{-1}\)
& one-site replacement
& \(1.99\times10^{-16}\)\\
RNA-like SCFG
& \(2.052488\times10^{-3}\)
& \(\mathcal{R}_{p,x}(x_p)=Z_0(x)\)
& \(0\)\\
RNA-like SCFG
& \(2.052488\times10^{-3}\)
& one-site recombination
& \(8.45\times10^{-16}\)\\
RNA-like SCFG
& \(2.052488\times10^{-3}\)
& two-site background recombination
& \(1.69\times10^{-15}\)\\
RNA-like SCFG
& \(2.052488\times10^{-3}\)
& mutation-order agreement
& \(1.69\times10^{-15}\)\\
\hline
\end{tabular}
\end{table}

The main text reports the corresponding rounded values. Small
platform-dependent changes in the last few floating-point digits are
expected because the residuals measure round-off after algebraically
equivalent calculations with different evaluation orders.

\subsection{Diagnostics for omitted finite terms}
\label{sec:supp-numerical-diagnostics}

The HMM fixed instance was also used to quantify two approximations that are
not exact finite identities.

First, omitting the mixed term from a two-site replacement gives the
normalized non-additivity measure
\begin{equation}
  r_{\mathrm{add}}
  =
  \frac{
    \left|
      \Delta_{p,q}Z
      -
      \left(
        \Delta_pZ+\Delta_qZ
      \right)
    \right|
  }{
    Z(x)
  }
  =
  \frac{
    \left|
      \Delta_p\Delta_q Z
    \right|
  }{
    Z(x)
  }.
  \label{eq:supp-numerical-radd}
\end{equation}
Its maximum in the fixed HMM instance was
\begin{equation}
  r_{\mathrm{add}}=0.9618,
  \label{eq:supp-numerical-radd-value}
\end{equation}
reported as \(0.96\) in the main text. This quantity is normalized by the
original partition function; it is not a relative error divided by the true
two-site mutation effect.

Second, for a one-site mutation, the exact log-partition change was compared
with the first-order transformation of the exact finite change in \(Z\):
\begin{equation}
  r_{\log}
  =
  \left|
    \Delta_p\log Z
    -
    \frac{\Delta_pZ}{Z(x)}
  \right|.
  \label{eq:supp-numerical-rlog}
\end{equation}
The maximum was
\begin{equation}
  r_{\log}=0.5788\ \text{nats},
  \label{eq:supp-numerical-rlog-value}
\end{equation}
reported as \(0.58\) nats in the main text. These diagnostics illustrate
that exact finite reconstruction of \(Z\) does not imply exactness of an
additive multisite approximation or of a first-order transformation of
\(\log Z\).

\subsection{Random-parameter sweep}
\label{sec:supp-numerical-random-sweep}

To check that the machine-precision agreement was not specific to one hand
chosen parameter set, the complete tests were repeated for 50 seeded random
parameter sets. These random distributions are synthetic numerical stress
tests and are not intended to represent standard biological alignment
scoring conventions. For each seed
\begin{equation}
  s=0,1,\ldots,49,
\end{equation}
the script initializes
\begin{equation}
  \texttt{rng = np.random.default\_rng(seed)}
\end{equation}
and then generates the HMM, SCFG, RNA-like SCFG, and alignment instances,
in that order, from the same random-number stream. The distributions are as
follows.

For the HMM,
\begin{equation}
  A_{ij},E_{ic},\pi_i
  \overset{\mathrm{ind}}{\sim}
  \mathrm{Uniform}(0.1,1.0).
  \label{eq:supp-numerical-random-hmm}
\end{equation}
Normalization is intentionally not imposed: the finite-replacement
identities require positive local factors, not stochastic normalization.

For the SCFG, the seven binary rules in
Eq.~\eqref{eq:supp-numerical-scfg-rules} retain the same support, with each
nonzero weight independently sampled from
\begin{equation}
  \mathrm{Uniform}(0.1,1.0),
\end{equation}
and all lexical factors are independently sampled from
\begin{equation}
  \mathrm{Uniform}(0.1,1.0).
  \label{eq:supp-numerical-random-scfg}
\end{equation}

For the RNA-like SCFG,
\begin{align}
  t_P,t_L,t_R
  &\overset{\mathrm{ind}}{\sim}
  \mathrm{Uniform}(0.1,0.5),
  \nonumber\\
  t_B
  &\sim
  \mathrm{Uniform}(0.05,0.3),
  \qquad
  t_E=1,
  \label{eq:supp-numerical-random-rna-rules}
\end{align}
the entries of \(e_L\) and \(e_R\) are independently sampled from
\(\mathrm{Uniform}(0.1,0.5)\), and each of the six allowed ordered pair
factors \(e_P(a,b)\) is independently sampled from
\(\mathrm{Uniform}(0.3,2.0)\). Disallowed pairs remain zero.

For the alignment model, every score-matrix entry is independently sampled
from
\begin{equation}
  \mathrm{Uniform}(-1.0,1.5),
\end{equation}
and
\begin{equation}
  d\sim\mathrm{Uniform}(1.0,3.0),
  \qquad
  e\sim\mathrm{Uniform}(0.2,1.0),
  \qquad
  \tau\sim\mathrm{Uniform}(0.5,1.5).
  \label{eq:supp-numerical-random-alignment}
\end{equation}

Table~\ref{tab:supp-numerical-random-results} gives the largest normalized
residual observed for each model over the 50-seed sweep in the reference
audit.

\begin{table}[t]
\centering
\small
\caption{Worst residuals over 50 seeded random parameter sets. The seed is
the seed of the shared \texttt{default\_rng} stream described above.}
\label{tab:supp-numerical-random-results}
\begin{tabular}{llll}
\hline
Model & Maximum residual & Seed & Identity at maximum\\
\hline
HMM
& \(5.607\times10^{-15}\)
& 35
& two-site\\
Affine-gap alignment
& \(2.052\times10^{-15}\)
& 32
& one-site\\
SCFG
& \(1.209\times10^{-15}\)
& 46
& one-site\\
RNA-like SCFG
& \(1.173\times10^{-14}\)
& 4
& two-site\\
\hline
\end{tabular}
\end{table}

Thus every residual in the 50-seed sweep was below
\begin{equation}
  2\times10^{-14},
\end{equation}
the bound stated in the main text. No NaN, infinity, or zero-division event
occurred in the audited run.

\subsection{Execution environment and reproducibility scope}
\label{sec:supp-numerical-environment}

The reference numerical audit used CPython~3.12.9 and NumPy~2.4.6 and ran
\begin{equation}
  \texttt{python3 verify\_numerics.py}.
\end{equation}
Two consecutive executions produced identical output, exited with status
zero, and produced no standard-error output or warnings. The same script was
also executed from a freshly extracted submission package.

The exact last digits of round-off residuals can depend slightly on the
Python/NumPy version, BLAS implementation, compiler, and hardware. The
scientifically relevant reproducibility targets are therefore (i) the fixed
partition functions, (ii) agreement of all exact identities at
double-precision round-off, (iii) the random-sweep bound
\(r_{\mathrm{exact}}<2\times10^{-14}\), and (iv) the finite-effect
diagnostics \(r_{\mathrm{add}}\simeq0.96\) and
\(r_{\log}\simeq0.58\).

Finally, the RNA numerical tests concern only the context-independent
five-rule RNA-like SCFG explicitly derived in Supplementary Section~S4.2.
They check its four-role one-site recombination and successive-background
two-site construction. They do not independently validate the full
dangling-end, terminal-mismatch, multiloop, and boundary-context case
enumeration of a Turner nearest-neighbor Rchange implementation. The latter
is the published algorithmic construction summarized separately in
Supplementary Section~S4.3.

\renewcommand{\refname}{Supplementary References}
\bibliographystyle{plainnat}
\bibliography{references}